\documentclass{article}

\usepackage[dvips]{color}

\usepackage{natbib}
\usepackage{setspace}
\usepackage{amsmath}
\usepackage{graphicx}
\usepackage{verbatim}
\usepackage{rotating}
\usepackage{multirow}
\usepackage{longtable}
\usepackage{booktabs}
\newcommand{\bigcell}[2]{\begin{tabular}{@{}#1@{}}#2\end{tabular}}
\usepackage{subfigure}
\usepackage{lineno}

\usepackage{fancyhdr}
\begin{document}


\title{Environmental and fishing effects on the dynamic of brown tiger prawn ({\it Penaeus esculentus}) \\ in Moreton Bay (Australia)}

\author{Marco Kienzle\footnote{DAFF biometry, Brisbane, Queensland, Australia}, Anthony J. Courtney\footnote{DAFF agri-science, Brisbane, Queensland, Australia.} and Michael F. O'Neill\textsuperscript{\dag}}

\maketitle

\begin{abstract}
This analysis of the variations of brown tiger prawn ({\it Penaeus esculentus}) catch in the Moreton Bay multispecies trawl fishery estimated catchability using a delay difference model. It integrated several factors responsible for variations in catchability: targeting of fishing effort, increasing fishing power and changing availability. An analysis of covariance was used to define fishing events targeted at brown tiger prawns. A general linear model estimated inter-annual variations of fishing power. Temperature induced changes in prawn behaviour played an important role in the dynamic of this fishery. Maximum likelihood estimates of targeted catchability ($3.92 \pm 0.40 \ 10^{-4}$ boat-days$^{-1}$) were twice as large as non-targeted catchability ($1.91 \pm 0.24 \ 10^{-4}$ boat-days$^{-1}$). The causes of recent decline in fishing effort in this fishery were discussed.
\end{abstract}




\linenumbers

\section{Introduction}

\begin{figure}[h!]
     \begin{center}
 \includegraphics[scale=0.5, angle = 0]{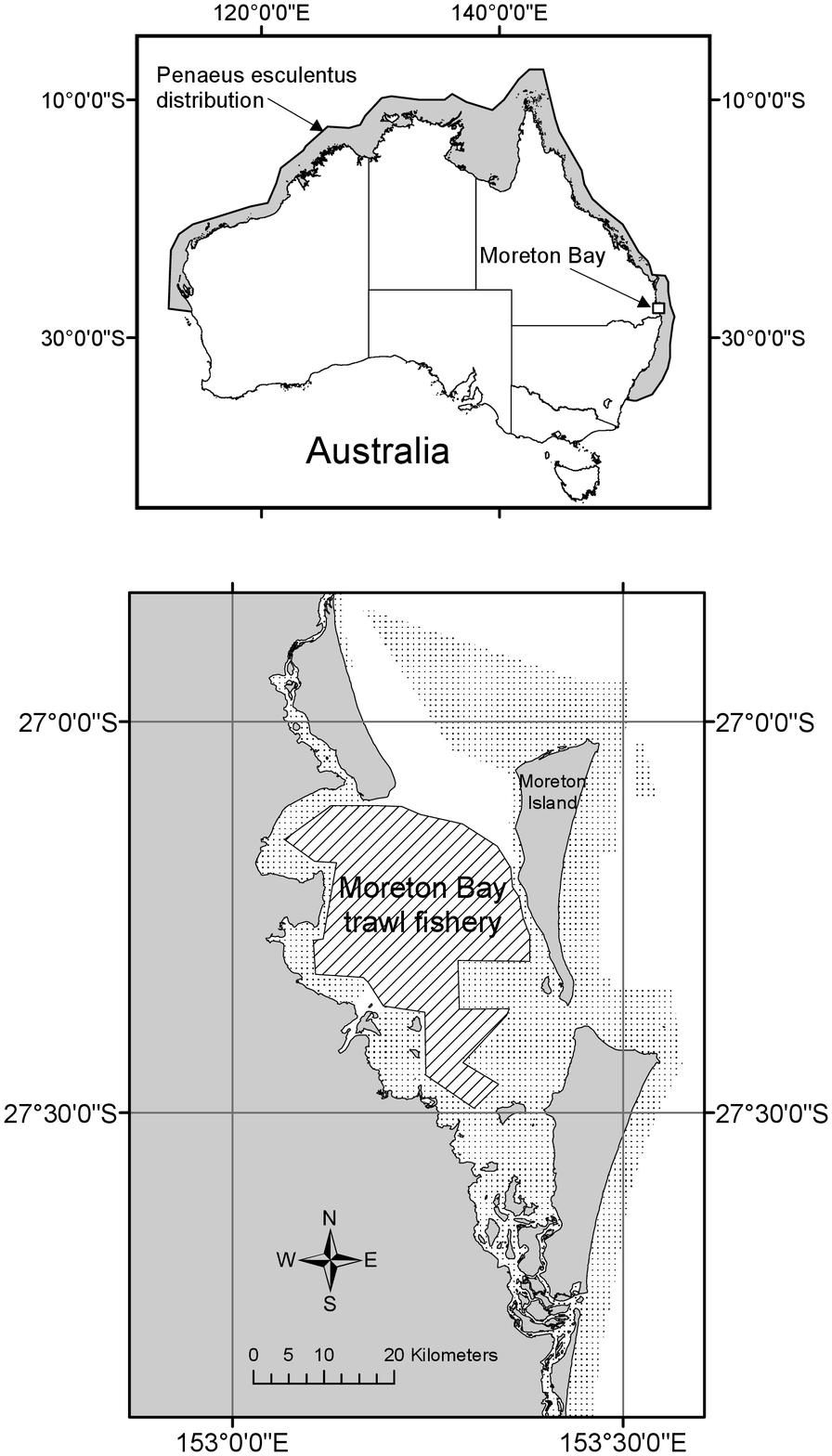}
\caption{}
       \label{fig:Map}
     \end{center}
  \end{figure}

\begin{figure}[!ht]
 \subfigure[]{ 
    \label{fig:TotalPrawnInfo:a}
    \begin{minipage}[b]{0.45\textwidth}
      \includegraphics[scale=0.28, angle = -90]{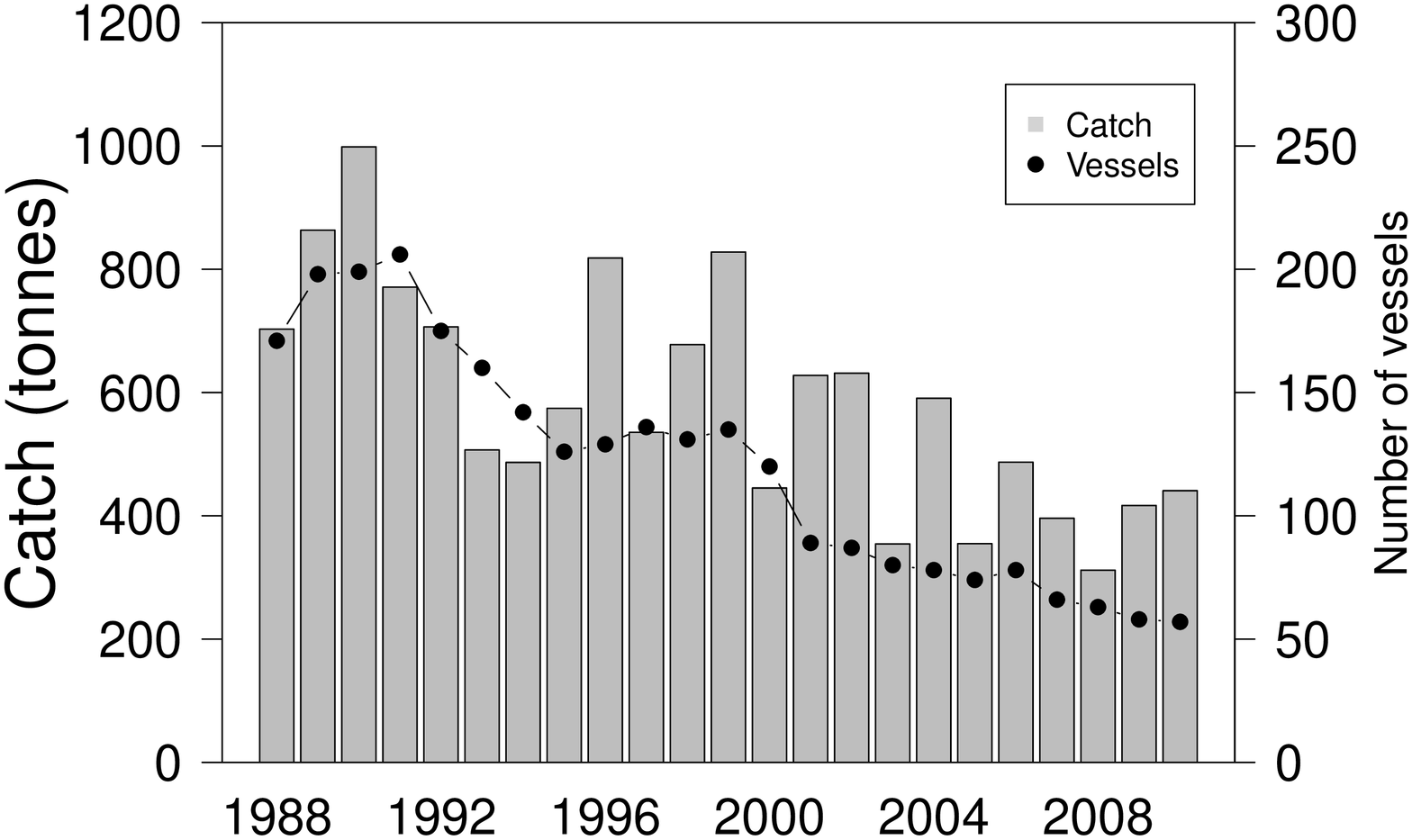}
   \end{minipage}
}
 \subfigure[]{ 
   \hspace{1cm}
    \label{fig:TotalPrawnInfo:b}
    \begin{minipage}[b]{0.45\textwidth}
    \includegraphics[scale=0.28, angle = -90]{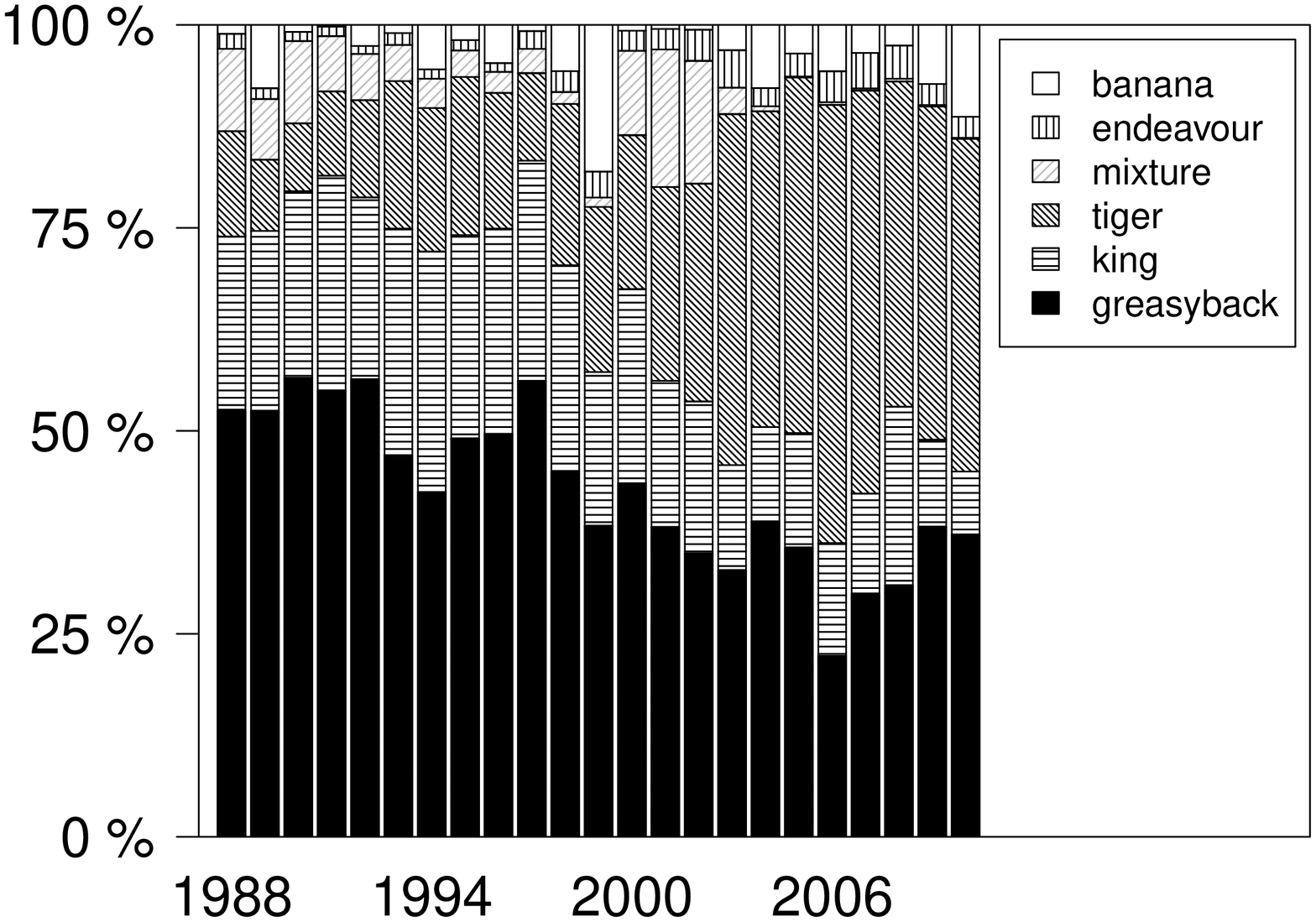}
   \end{minipage}
}
\caption{}
    \label{fig:TotalPrawnInfo}
  \end{figure}

Moreton Bay covers a 1500 km$^{2}$ area on the east coast of Australia (Fig.~\ref{fig:Map}). Its shallow water ($< 36$ m, average 6.8 m) provide habitats suitable for at least 12 species of prawn \citep{Hyland87PhD}, five of which (greasyback ({\it Metapenaeus bennettae}), eastern king ({\it Melicertus plebejus}), brown tiger ({\it Penaeus esculentus}), endeavour ({\it Metapenaeus endeavouri}) and banana prawns ({\it Fenneropenaeus merguiensis})) are caught by a commercial otter trawl fishery. This industry grew rapidly after 1952 prompting the government to regulate the expansion of the fleet \citep{parke2013b}. Today, the fishery is managed by input controls in the form of limited entry, vessel and gear restrictions. Spatial closures were introduced in 1993 by the creation of a Marine Park which coverage was extended from 0.5\% to 16\% of the Bay in 2009. The fishery has provided predominantly small prawns (less than 20 g) to the local market for human consumption and bait. Total catch in 1952--53 was 136 tonnes (t) and reached up to a 1000 t in 1990. In recent years, the total number of vessels in the fishery has declined by 70\%, from a peak at 206 vessels to 57 (Fig.~\ref{fig:TotalPrawnInfo:a}, \citep{SeaCRC2012}) in response to falling demand, decreasing prawn prices and increasing fuel costs \citep{Pascoe2013115}. Landings's species composition shifted from being dominated by small size prawns, mostly greasybacks and juveniles eastern king, to an increasing proportion of larger, more valuable, brown tiger prawn (Fig.~\ref{fig:TotalPrawnInfo:b}).\\

The brown tiger prawn is endemic to Australia and distributed across tropical and subtropical coastal waters (Fig.~\ref{fig:Map}) in depths to 200 m. A characteristic of penaeid coastal shrimps is their short life span, of the order of two years, and their presence in the fishery in significant quantities for a period generally little more than a year \citep{garcia1981life}. In Moreton Bay, the population of brown tiger prawns has been assumed to have largely non overlapping generations \citep{MEC:MEC3132} because previous estimates of natural and fishing mortality \citep{Wang99a, som97a} suggested that less than 1\% of a cohort would be alive after 1 year of exploitation. Most eggs are produced in a single, clearly defined peak in October--November although spawning continues to May each year \citep{court97a}. Larval survival depends on water temperature and salinity. Nursery habitats for {\it P. esculentus} are shallow inshore areas prone to estuarine fluctuations of temperature and salinity \citep{Keys2003325}. Peak settlement of juveniles in sea grass in the southern part of the Bay occurs between September and November and late January and April \citep{OBrien94a}. Brown tiger prawns recruit to the fishery at a large size (20 g and 27 mm carapace length) compared to the other species \citep{court95a}. Adult brown tiger prawns are benthic and nocturnal, remaining buried during the day and emerging in the evening to feed and mate \citep{Keys2003325}. The duration of nightly emergence from the substrate and the rate of activities such as swimming and foraging are dependent on diurnal rhythms and water temperature \citep{hil85a}, light intensity and moulting events \citep{Keys2003325}. Catchability of {\it P. esculentus} in wild fisheries has been linked with temperatures \citep{whit75a} and was postulated to play an important role in determining the magnitude of brown tiger catch in Moreton Bay \citep{hil85a}.\\

The Moreton Bay fishery is a small component of Queensland's East Coast Otter Trawl fishery (ECOTF, \citep{Pascoe2013115}) which fishing capacity increased significantly since the introduction of otter trawling in this region. New technologies effective at improving fishing efficiency were quickly adopted by this fleet \citep{rob98a}. Since the mid-1980s, fishing power grew between 0.5 and 4.7\% yr$^{-1}$ depending on the sub-fishery considered \citep{bis08a, ONeill2006R, ONeil2007a}. Technology creep was always perceived as a concealed threat to their sustainability. As a consequence they have been closely monitored and subject to quantitative stock assessments for many years in order to manage the risk of recruitment overfishing characteristic of tiger prawns \citep{Dichmont2006204}. Recent changes in the Moreton Bay trawl fishery have prompted the industry to investigate alternative fishing strategies but a lack of quantitative stock assessment for this area precluded such evaluation. Therefore a delay difference model \citep{sch85a} was applied to fill this gap. The present stock assessment was implemented to estimate catchability in order to quantify the impact of fishing on the survival of this population of brown tiger prawn. The model was developed to take into account the interactions between environment, technology and fishing effort. A major challenge arose from applying this single species stock assessment model to data from this multispecies fishery and required to define the portion of effort targeted at the brown tiger prawns. Model uncertainties on several aspects of the fishery generated a large amount of plausible stock assessment models that were fitted to the data by maximum likelihood \citep{Burnb03} to identify which hypotheses best described historical variations of brown tiger catch in Moreton Bay .

\section{Materials and methods} 

   \subsection{Data sources}

   The present assessment of the Moreton Bay stock of brown tiger prawns is based on (a) compulsory commercial logbook data collected since 1988 that provided total catch by species, vessel and day and (b) skipper interviews conducted in 2000 and 2010 used to construct the history of fishing gear used by each vessel. The fishing power analysis used this information at the highest temporal resolution (catch by vessel per day). The stock assessment model was fitted to weekly catches grouped into 22 biological years (1989--2010). Each biological year was made of 52 weeks, started around the first of July when brown tiger prawn activity in Moreton Bay was at its lowest and finishing a year later at the end of June.\\

Logbook records from 1988 to 2010 were combined with vessel and gear descriptions collected during the skipper interviews. The number of nets, total head-rope length, mesh size, type and size of the otter-boards, steaming speed, engine power, propeller diameter, presence/absence of kort nozzle, maximum trawling speed and engine revolution speed were combined using the Prawn Trawl Prediction Model \citep{sterlingPHD, bis08a} into an estimate of swept area rate (SAR, in hectares per hour) for each vessel/net configuration available (134 in total). Ten fishing technologies (colour echo-sounder, satellite navigation, global positioning system (GPS), plotter, auto-pilot, GPS coupled with auto-pilot, by-catch-reduction device (BRD) and turtle excluding devices (TED)) were coded as binary variables to indicate presence or absence on-board a vessel during each fishing event. Finally, a continuous variable describing the moon phase was also associated with each logbook record.\\

\begin{figure}[!ht]
 \subfigure[]{ 
    \label{fig:a}
    \begin{minipage}[b]{0.45\textwidth}
    \includegraphics[scale=0.25, angle = -90]{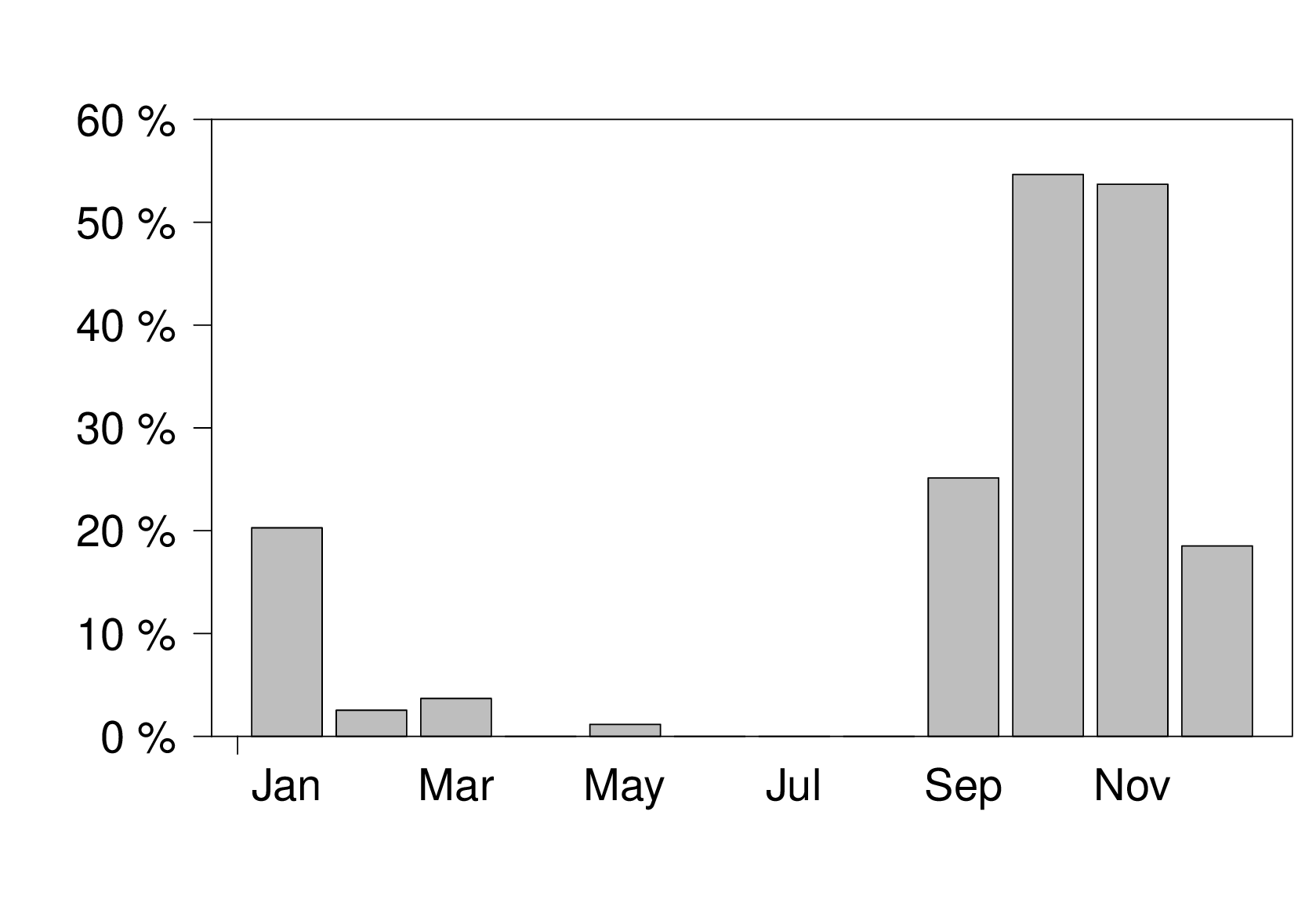}
   \end{minipage}
}
 \subfigure[]{ 
    \label{fig:b}
    \begin{minipage}[b]{0.45\textwidth}
    \includegraphics[scale=0.25, angle = -90]{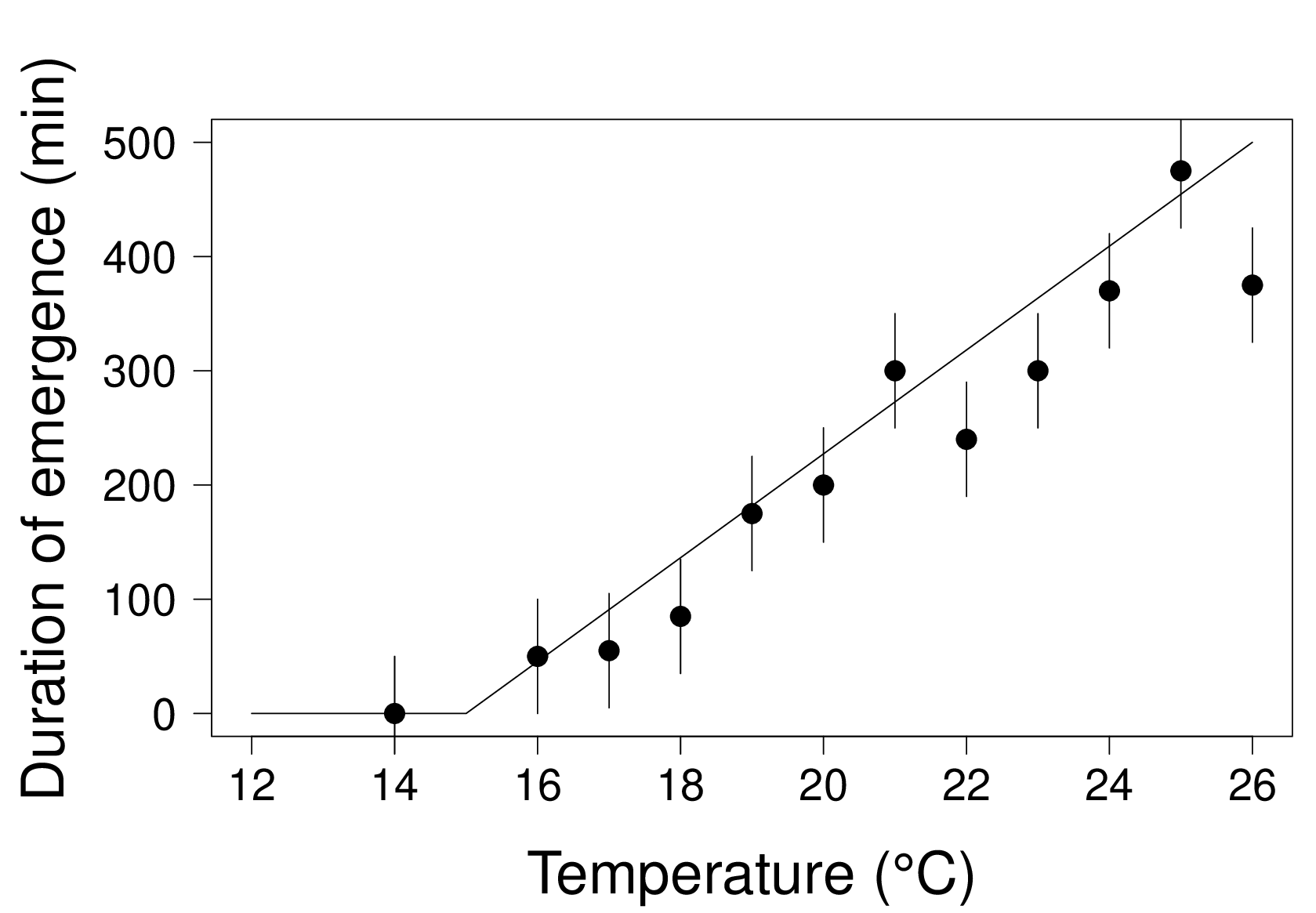}
   \end{minipage}
}
 \subfigure[]{ 
    \label{fig:c}
    \begin{minipage}[b]{0.45\textwidth}
    \includegraphics[scale=0.25, angle = -90]{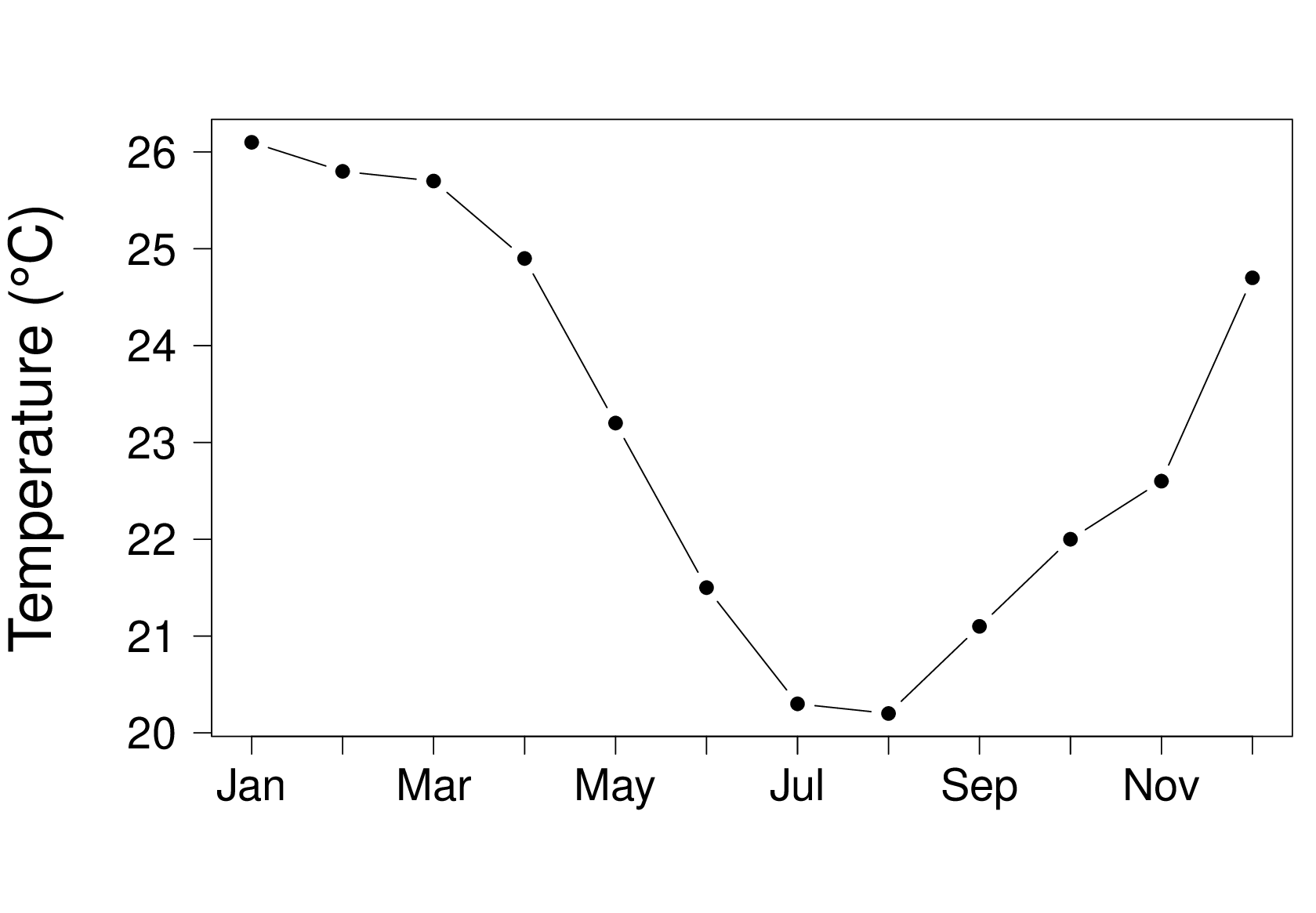}
   \end{minipage}
}
 \subfigure[]{ 
    \label{fig:d}
    \begin{minipage}[b]{0.45\textwidth}
    \includegraphics[scale=0.25, angle = -90]{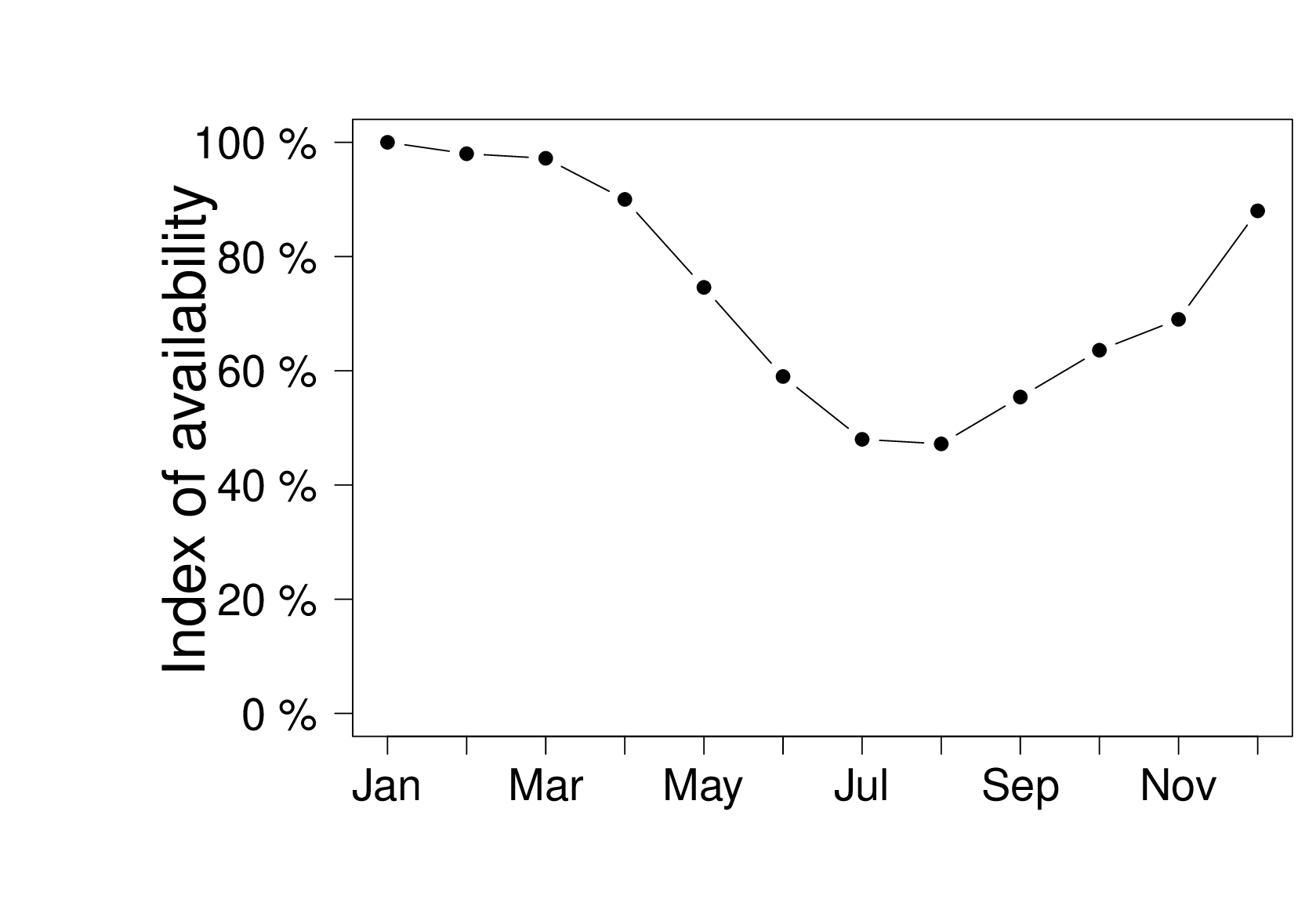}
   \end{minipage}
}
\caption{}
    \label{fig:EstimatesOfAvailability}
  \end{figure}

Sea surface temperatures in Moreton Bay vary from 16$^{o}$C in winter to 29$^{o}$C in summer whereas its range at Cape Moreton is attenuated to 18.5--25.5$^{o}$C by the ocean \citep{nla.cat-vn5267095}. Seasonal average sea surface temperatures collected within a 60 nautical miles radius around Moreton Island (\cite{metoc2012}, Fig.~\ref{fig:c}) were combined with experimental duration of emergence of tiger prawns from \cite{hil85a} (Fig.~\ref{fig:b}) to create a seasonal index of brown tiger prawn availability ($\gamma$, Fig.~\ref{fig:d}). This index was made to vary weekly from high availability in summer to low availability in winter. Its amplitude represented a decline in availability of about 50\% between summer and winter. This variable was kept constant between years. It was included in several versions of the delay difference model as a multiplier of catchability to determine if the hypothesis that variations of temperature influence the magnitude of brown tiger prawn catch in Moreton Bay \citep{hil85a} was supported by logbooks data. \\

   \subsection{Fishing power analysis}

\begin{figure}[h!]
     \begin{center}
       \includegraphics[scale=0.5, angle = -90]{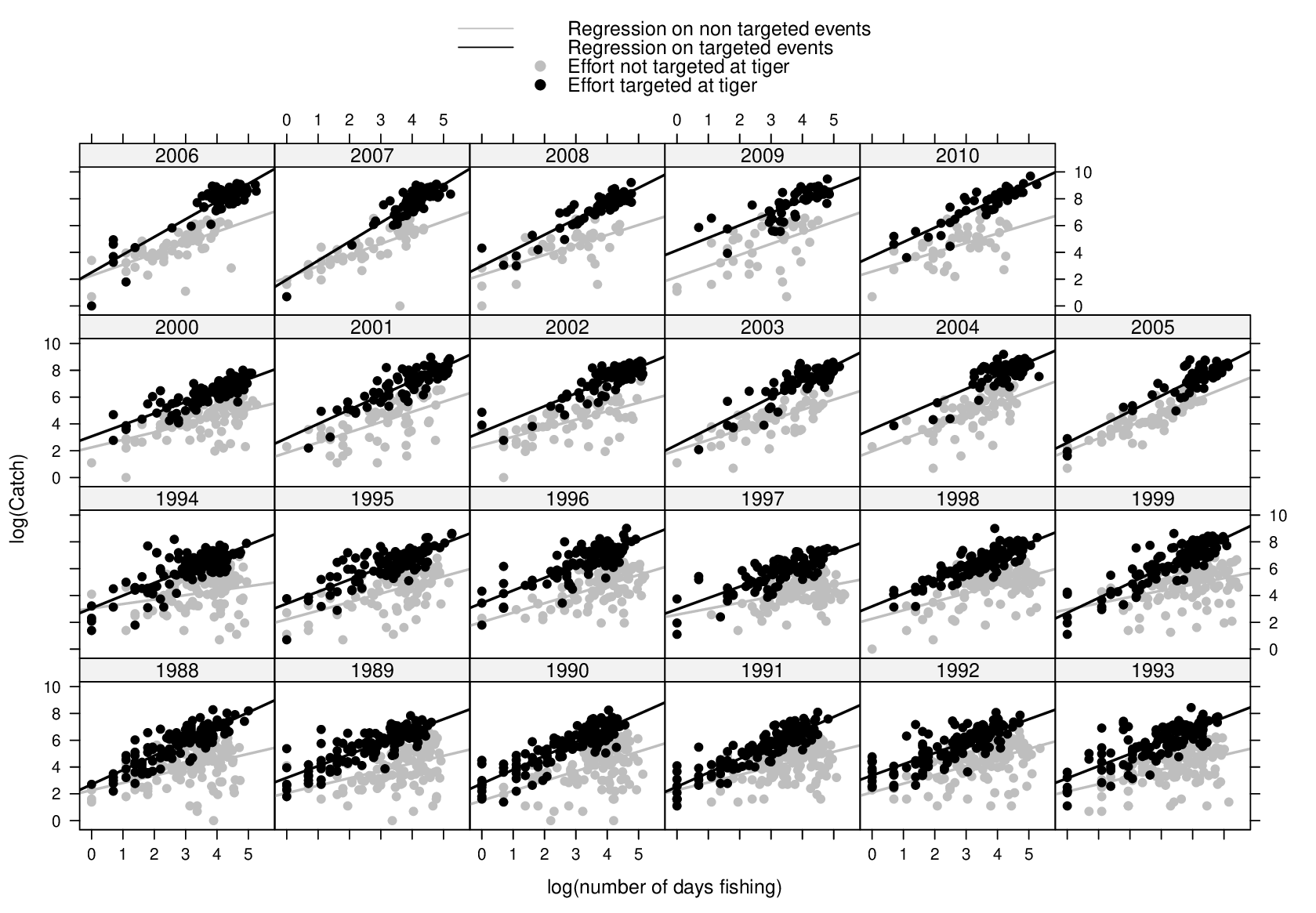}
       \caption{}
       \label{fig:TargetvsNonTarget}
     \end{center}
   \end{figure}

In fisheries, catch is often found to increase linearly as a function of effort on the log-scale \citep{hil92b}. Multispecies fisheries such as the Moreton Bay fishery exploit different species opportunistically throughout the year as they become available. An un-discriminated analysis of the data showed no relationship between brown tiger catch and fishing effort. This was the result of including records with very low catch rates across the range of effort because non-target species were caught at random or were present on the fishing ground at a lower abundance than the target species (Fig.~\ref{fig:TargetvsNonTarget}).
A rule to classify each unit of fishing effort (in boat-days) into fishing targeted or not targeted at brown tiger prawn was required to analyze these data. A large number of targeting rules were proposed and assessed against the data using an analysis of covariance (ANCOVA, Tab.~\ref{tab:ANCOVA}). The data were partitioned according to each rule and both targeted and non-targeted groups of data were fitted with a separate linear regression between catch and effort on the log-scale (Fig.~\ref{fig:TargetvsNonTarget}). The minimum residual sum of square of an ANCOVA was used to choose the targeting rule that explained the largest variability in the data. \\

Catch and effort data from fishing records classified as targeted at brown tiger prawns were standardized using a Generalized Linear Model (GLM) to estimate yearly variation of fishing power \citep{maun04a}. Logarithm of daily brown tiger catch for each vessel (C$_{i}$ in kg) was expressed as a linear combination of (a) the logarithm of area swept (SA) by the otter trawl estimated as the product of SAR and number of hours fished; (b) several binary variables coding for presence or absence of particular technologies $j$ (represented by matrix X$_{i,j}$); (c) an index of abundance for years, months (taken as factors) and their interaction (represented by matrix Y$_{i,y,m,l}$) and (d) lunar phase (L$_{i}$)

\begin{equation}
{\rm log}({\rm C}_{i}) = a + b \ {\rm log}({\rm SA}_{i}) + {\rm X}_{i,j} \ c_{j} + {\rm Y}_{i,y,m,l} \ d_{k,l,m} + e \ {\rm L}_{i}
\end{equation}

The parameters of the linear model were estimated in R \citep{R} using a Generalized Linear Model (GLM) with quasi family, log-link and variance proportional to the square of mean. An alternative fit using the Gaussian family with log-link provided a poorer fit to the data and was abandoned.\\

Multicollinearity in the data set was identified and treated regressing all pairs of explanatory variables against each other \citep{drapb} to calculate the variability ($R^{2}$) of one variable that was explained by the other (Tab.~\ref{tab:MultiCollinearity-LMonPairsOfVariable}); one variable in each pair was discarded from the pool of possible GLM co-variates when R$^{2}$ \textgreater \hspace{0.1cm} 20\%. The same procedure using a threshold value of 5\% eliminated many more variables from the GLM and was abandoned because it was deemed too stringent. This approach to treat multicollinearity provided stability to the parameter estimates, in particular to the estimates of fishing power.\\

An estimate of fishing power variations between 1988 and 2010 was obtained using the GLM to calculate the average catch per hour trawled using a fixed level of abundance \citep{bis08a}. These estimates were expressed relative to the beginning of the time-series (1988) were incensitive to the fixed level of abundance chosen. Standard errors for the relative fishing power time-series were obtained by propagating uncertainties from the GLM predictions \citep{bevrob03}.\\

   \subsection{Stock assessment model}

\subsubsection{Population dynamics}

A Schnute-Deriso delay-difference model \citep{sch85a, Der80a, hil92b} was used to estimate weekly variations in biomass ($B_{t}$) of brown tiger prawns in Moreton Bay between 1989 and 2010 

\begin{equation}
B_{t} = s_{t-1} \ B_{t-1} + \rho \ s_{t-1} \ B_{t-1} - \rho \ s_{t-1} \ s_{t-2} \ B_{t-2} - s_{t-1} \ \rho \ w_{k-1} \ R_{t-1} + w_{k} \ R_{t} \ ,\  3 \leq t \leq 264
\end{equation}

Sex-combined growth parameters ($\rho$, $w_{k-1}$ and $w_{k}$), derived from von Bertalanffy estimates \citep{grib94a}, were fixed in the model (Tab.~\ref{tab:ModelPar}). This model assumed all prawns were fully recruited to the fishery (knife-edged selectivity) at an age of 22 weeks ($k=22$), weighing 19.5 grams. $w_{k-1}$, the pre-recruitment weight was interpreted as a parameter rather than the actual weight at age $k-1$ and was estimated according to \cite{sch85a}. Survival ($s_{t}$) varied as a function of a fixed natural mortality rate ($M$, Tab.~\ref{tab:ModelPar}) and fishing mortality ($F_{t}$) proportional to effort ($F_{t} = q E_{t}$)

\begin{equation}
s_{t} = \exp[- (M + q E_{t})]
\end{equation}

\noindent where catchability ($q$) was estimated.\\

This model estimated the magnitude of recruitment in each week ($R_{t}$) using 1 parameter to describe total recruitment in each year ($R_{y}$) between 1989 and 2010 and 2 parameters ($\mu$, $\kappa$) from the von Mises probability density function \citep{mardia1999directional} to allocate a proportion of the total recruitment within each year to each 52 weeks

\begin{equation}
f(x|\mu, \kappa) = \frac{\exp[\kappa \ cos(x-\mu)]}{2 \pi I_{0}(\kappa)}
\end{equation}

\noindent where $I_{0}(x)$ is the modified Bessel function of order 0.\\

\subsubsection{Alternative fishing mortality models}

\begin{figure}[h!]
     \subfigure[]{ 
       \label{fig:FishPowerTrends}
    \begin{minipage}[b]{0.5\textwidth}
       \includegraphics[scale=0.25, angle = -90]{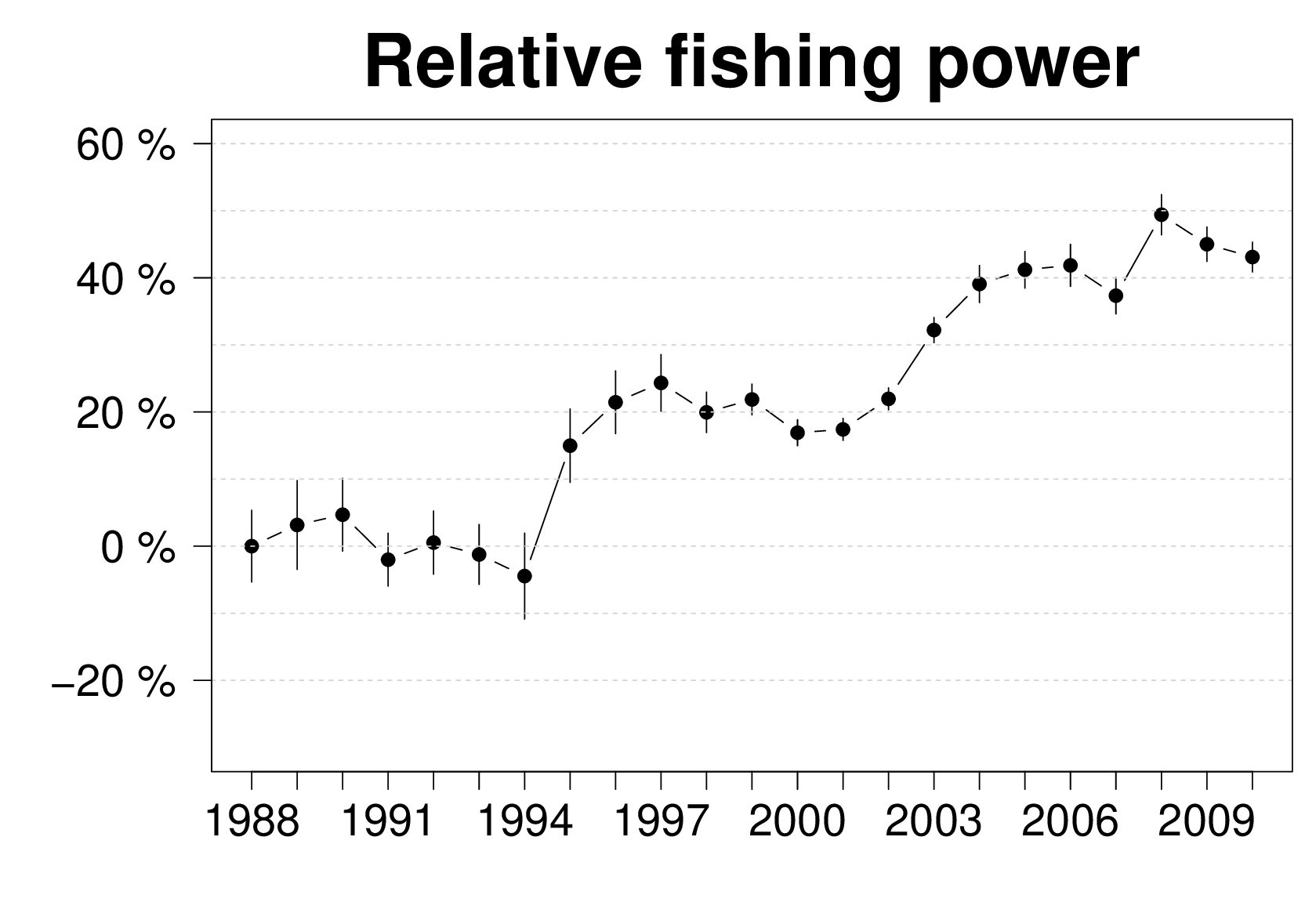}
\end{minipage}
}
     \subfigure[]{ 
       \label{fig:StandardizedEffort}
    \begin{minipage}[b]{0.5\textwidth}
       \includegraphics[scale=0.25, angle = -90]{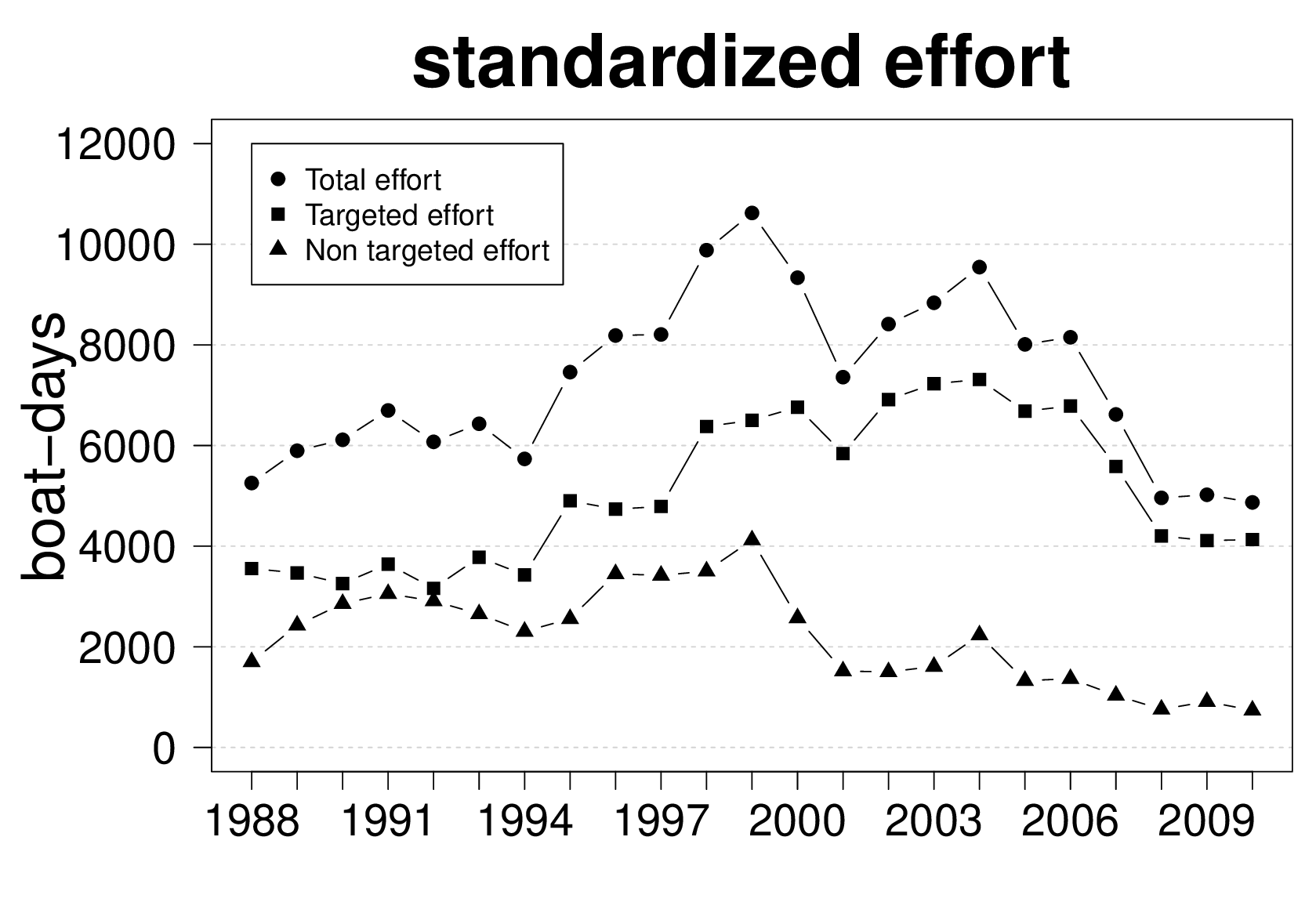}
\end{minipage}
}
\caption{}
       \label{fig:FishPower}
  \end{figure}

Effort was used as a co-variate in the delay difference model to compare the capacity of different fishing mortality models to explain the weekly variation in brown tiger prawn catch. A total of 7 models were evaluated: the first model used all effort reported catching tiger prawns (model 1 in Tab.~\ref{Tab:FishingMortalityModels}); model 2 used effort split into targeted and non-targeted effort to estimate two coefficients of catchability; the third and fifth models corrected time series of effort by yearly variations of fishing power (Fig.~\ref{fig:FishPowerTrends}); model 4 and 6 allowed for availability to vary within years as a function of temperature (Fig.~\ref{fig:EstimatesOfAvailability}). The seventh model was similar to model 4 but non-targeted effort was not corrected by fishing power.

\subsubsection{Fitting method}

A total of 28--29 parameters (1 or 2 catchability parameters, two initial biomass ($B_{1}$ and $B_{2}$), two von Mises parameters and 22 annual recruitment parameters and $\sigma$, the standard deviation of observation errors) were estimated by maximum likelihood assuming the square-root of predicted catch ($\hat{C}_{t}$, \citep{quin99b}) 

\begin{equation}
        \hat{C}_{t} = \frac{F_{t}}{ M + F_{t}}  B_{t} ( 1 - \exp[-(M + F_{t})])
\end{equation}

\noindent provided the mean of a Normal distribution of the square-root of observed catches ($C_{t}$) \citep{dichmont2003application} with residual standard deviation ($\sigma$). The negative log-likelihood function used to fit the model was \citep{haddon2010modelling}

\begin{equation}
-\log(L) = n \ log(\sqrt{2 \pi} \sigma) + \frac{1}{2 \sigma^{2}} \sum_{t=1}^{n=1144} \bigl ( \sqrt{\hat{C}_{t}} - \sqrt{C_{t}} \ \bigr ) ^{2}
\end{equation}

The logarithm was tested as an alternative transformation to the catch data but was abandoned due to non-normal errors. The model was implemented in C++ and used MINUIT minimization library \citep{minuit2} available through ROOT \citep{root}. Requests regarding the availability of the code should be directed to the first author of this publication, M. Kienzle. 

\subsubsection{Projections, equilibrium and reference points}

The proportion of mature female biomass estimated from scientific surveys (Fig~\ref{fig:a}, \cite{court97a}) were combined with the estimated stock biomass to calculate spawning stock biomass (SSB) assuming an even sex-ratio. A Ricker model was fitted to SSB and recruitment estimates lagged by 1 year using linear regression on transformed data \citep{hil92b}. The parameters of this stock-recruitment relationship and uncertainty were used to close the biological cycle by simulating recruitment on the log-scale using a Gaussian random number generator. The dynamic of the stock at several constant level of exploitation was projected over a 150-year period to calculate maximum sustainable yield (MSY) and associated effort (E$_{{\rm MSY}}$). In these simulations, effort was distributed within each year according to the average intra-annual pattern observed between 2006 and 2010. 

\section{Results} 

   \subsection{Fishery statistics}

\begin{figure}[h!]
     \begin{center}
       \includegraphics[scale=0.5, angle = -90]{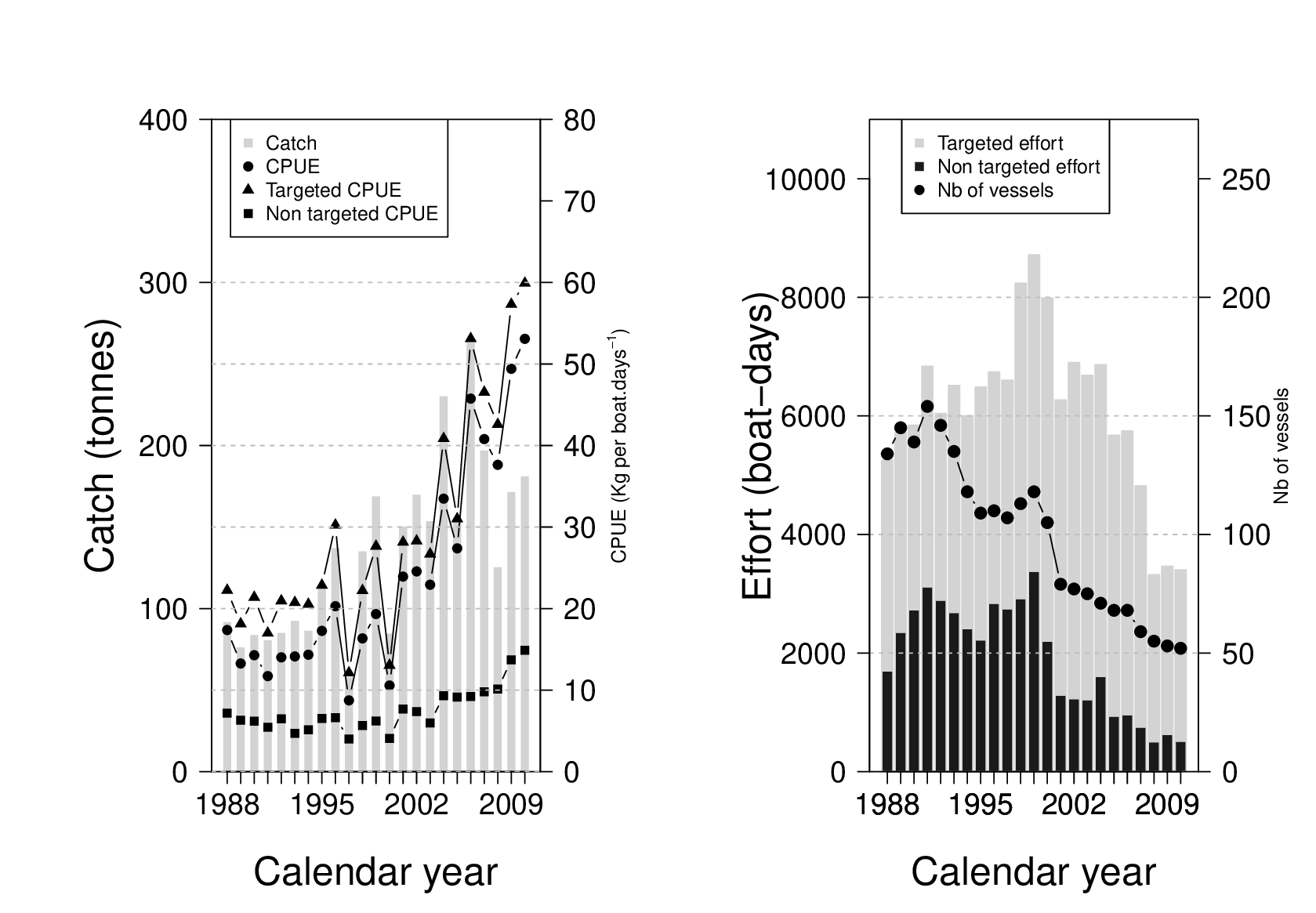}
\caption{}
       \label{fig:TSCatchEffortAndCPUE}
     \end{center}
  \end{figure}

Since the introduction of compulsory logbooks in this fishery in 1988, total catch of all prawn species nearly halved from a 5-year average of 743 $\pm$ 107 t at the beginning of the time series to 392 $\pm$ 66 t in the last 5 years (Fig.~\ref{fig:TotalPrawnInfo:a}). Over the same period, the total number of boats operating in Moreton Bay declined by 2/3 from a median of 198 to 66. Meanwhile, brown tiger prawn catches almost doubled, from an average of 100 tonnes per year before 2000 to an average of 180 tonnes afterwards (Fig.~\ref{fig:TSCatchEffortAndCPUE}). Nominal CPUE fluctuated around 16.0 $\pm$ 4.5  kg/boat-day until 2002 before increasing sharply to an average 40.2 $\pm$ 8.5 kg/boat-day between 2006 and 2010.

   \subsection{Effort targeted at brown tiger prawn}

Several rules were applied to classify daily individual vessel logbook records into targeted and non-targeted effort at brown tiger prawns. The residual sum of square of an ANCOVA reached a minimum when effort was assumed targeted at brown tiger if this species represented more than 20\% of the catch (Tab.~\ref{tab:ANCOVA} and Fig.~\ref{fig:TargetvsNonTarget}). According to this rule, 90\% of tiger catch was associated with targeted effort. Total effort reported catching brown tiger prawn increased from below 6000 boat-days in 1988 up to above 8000 boat-days in 1998--2000 and then declined to around 3500 boat-days (Fig.~\ref{fig:TSCatchEffortAndCPUE}). The proportion of non-targeted effort that was frequently above 30\% prior to 2000, declined to around 15\% in recent years.

   \subsection{Fishing power analysis}

Vessel identifier ({\it i.e.} Boat Mark) was confounded with most other co-variates (R$^{2} \geq \ 0.34$) and was eliminated from subsequent analysis (Tab.~\ref{tab:MultiCollinearity-LMonPairsOfVariable}). Auto-pilot was eliminated because it explained 42\% of the variability of colour echo-sounder. As a result of this selection process, catch data were fitted with a GLM using the following co-variates: year, month and their interaction; the logarithm of swept area (SA); lunar phase and the presence/absence of satellite navigation system (satnav); differential GPS (dGPS); plotter; GPS coupled with autopilot; computer mapping devices; by-catch reduction devices (BRD); turtle excluding devices (TED) and colour echo-sounder.\\

Single term deletion from the full model (performed using the {\it drop1} function in \cite{R}) indicated that abundance terms (year, month and interaction between year and month) explained the largest portion of catch variability (Tab.~\ref{tab:Drop1GLM}), followed by,  in decreasing order of importance, swept area (SA), computer mapping, satnav, plotter, BRD, lunar phase, dGPS, colour echo-sounder, GPS coupled with autopilot and TED. The satellite navigation systems were found to have a positive effect on fishing efficiency: satnav was estimated to improve catch by 25\%; dGPS by 7\% but effect of GPS was not found to be significant (Tab.~\ref{tab:FishPowerParEst}). Electronic mapping systems were also found to improve catch: computer mapping improved catch by 16\% and plotter by 12\% respectively. Bycatch reduction devices were found to improve catch of brown tiger by around 10\%. On the other hand, turtle excluding devices reduced catch by around 9\%. Fishing at full moon was less effective than at new moon. Autopilot coupled with GPS was found to reduce catch rates. \\

The combined effect of all fishing variables accounted in the GLM estimated fishing power increased by 40--50\% from 1988 to 2010 (Fig.~\ref{fig:FishPowerTrends}). Little fishing power variation could be identified between 1988 and 1994. Two large improvements occurred starting in 1994 and 2001, each enhancing the fishing efficiency of this fleet by 20\%. Uncertainties associated with this time-series declined through time as the number of skippers providing relevant information for this analysis increased.\\

   \subsection{Stock assessment}

\begin{figure}[h!]
     \subfigure[]{ 
       \label{fig:ModelAndData}
       \begin{minipage}[b]{0.95\textwidth}
       \includegraphics[scale=0.5, angle = -90]{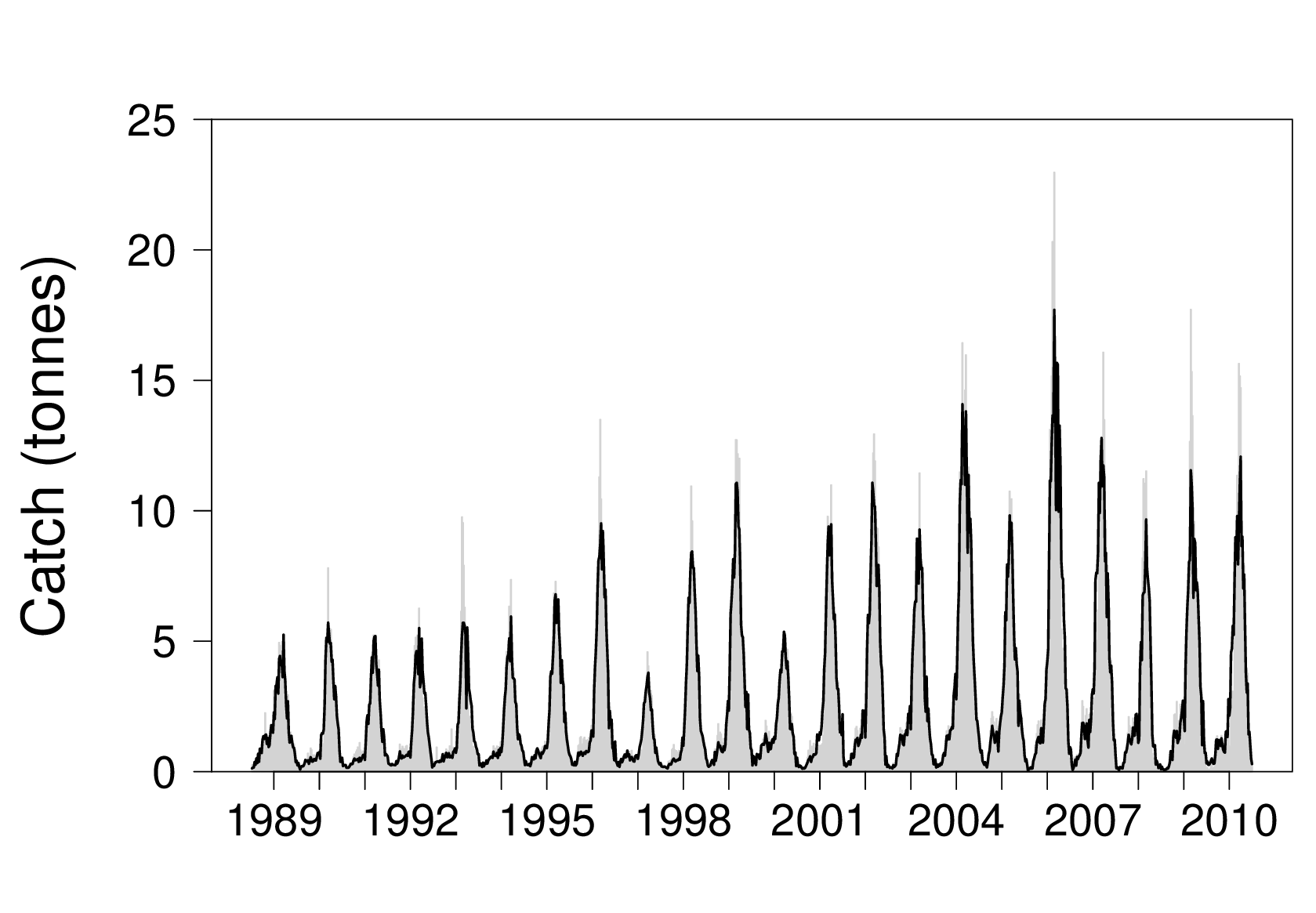}
       \end{minipage}
     }
\subfigure[]{ 
       \label{fig:ModelResiduals}
       \begin{minipage}[b]{0.5\textwidth}
       \includegraphics[scale=0.25, angle = -90]{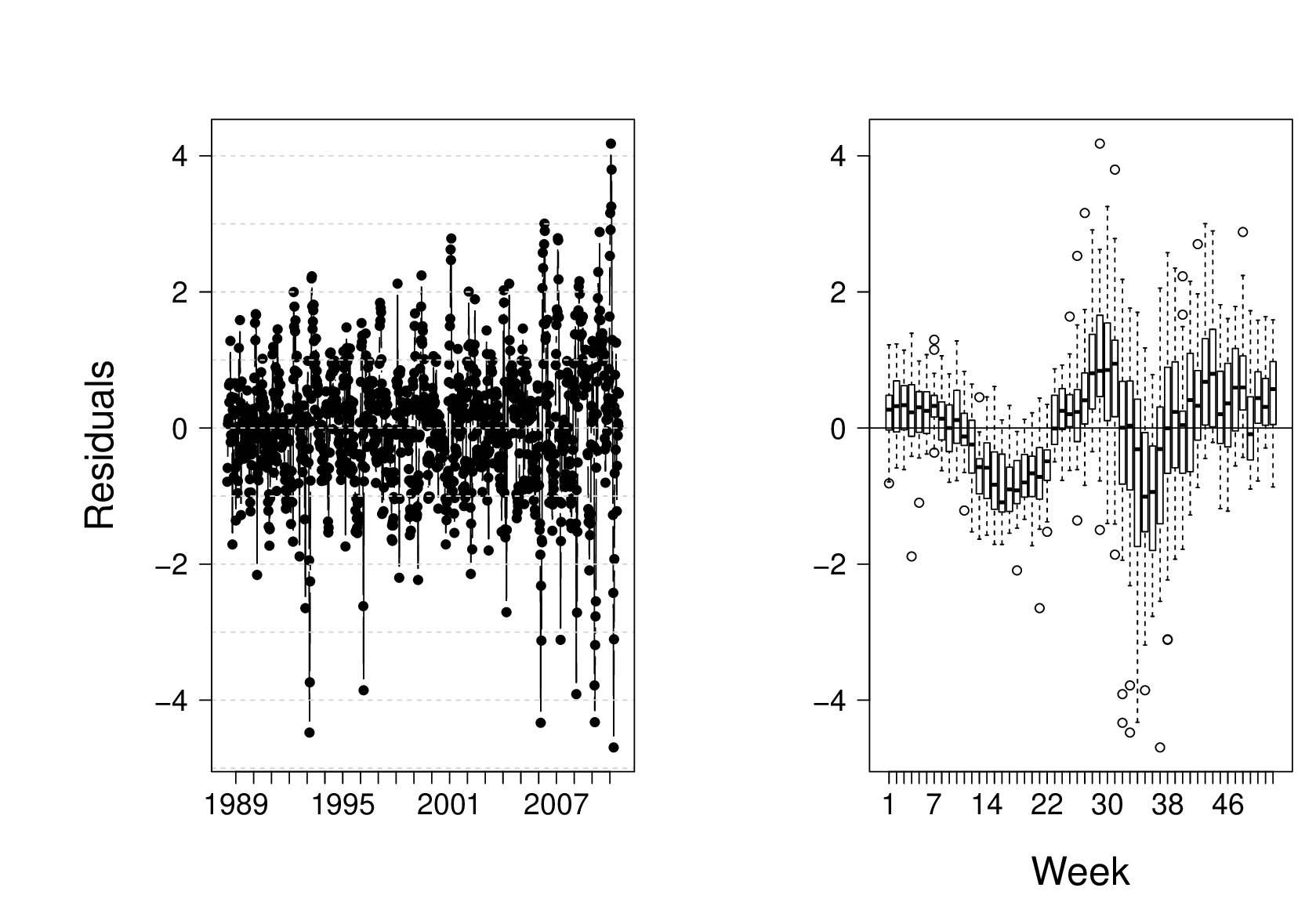}
       \end{minipage}
     }
     \subfigure[]{ 
       \label{fig:SeasonalDistOfRec}
       \begin{minipage}[b]{0.5\textwidth}
       \includegraphics[scale=0.25, angle = -90]{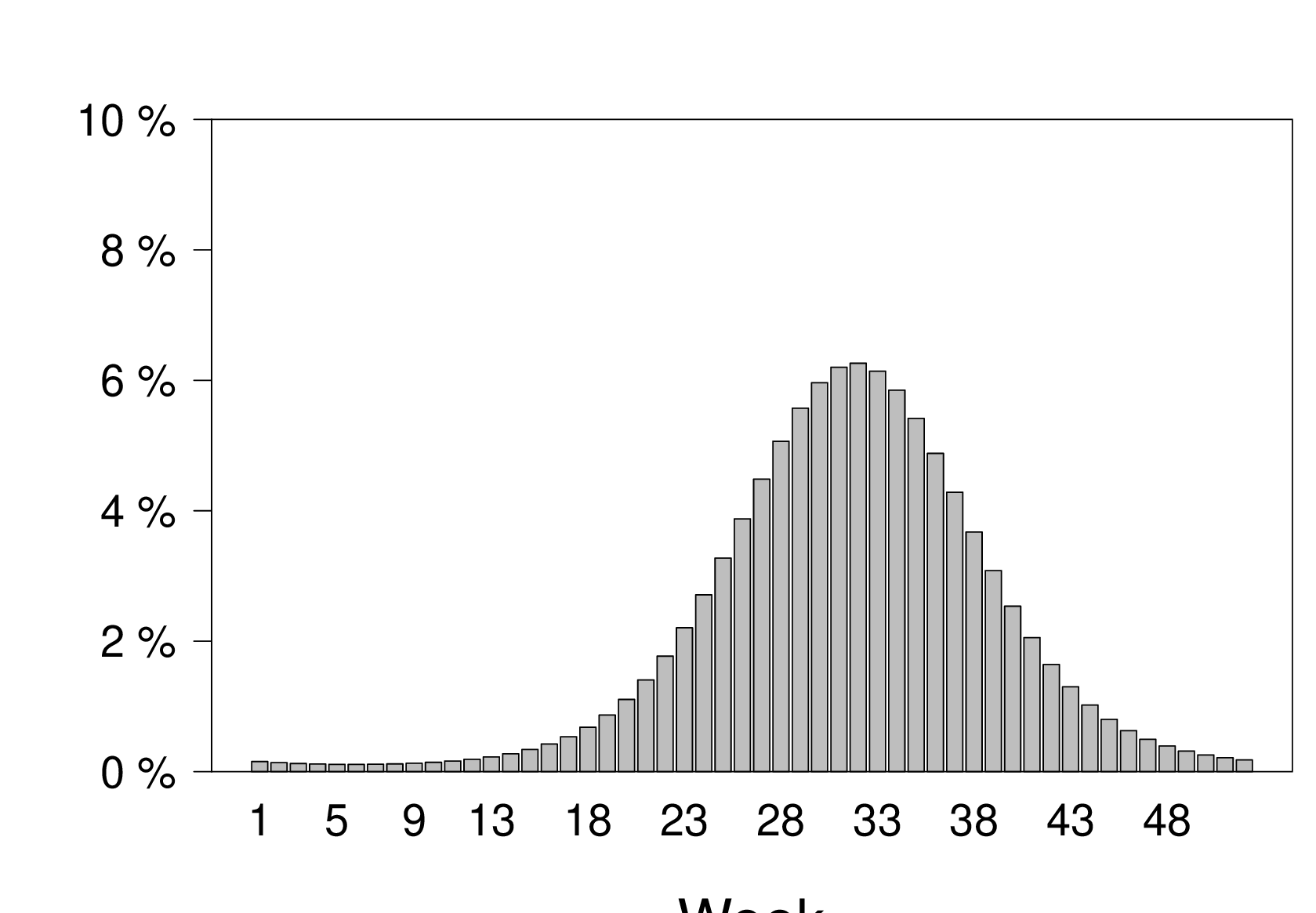}
       \end{minipage}
     }
     \subfigure[]{ 
       \label{fig:TSrecruitment}
       \begin{minipage}[b]{0.5\textwidth}
       \includegraphics[scale=0.25, angle = -90]{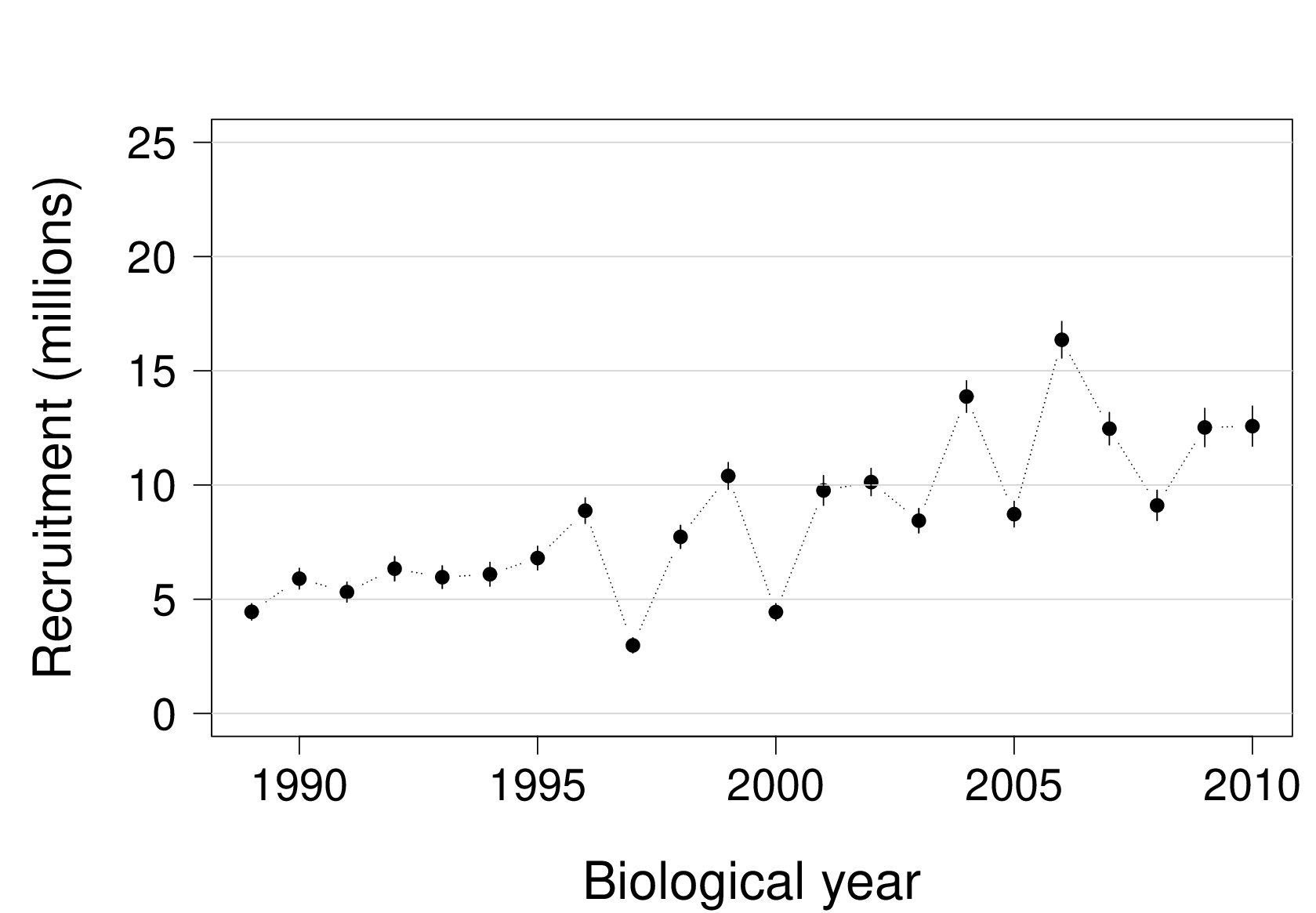}
       \end{minipage}
     }
     \subfigure[]{ 
       \label{fig:Ricker:a}
       \begin{minipage}[b]{0.5\textwidth}
       \includegraphics[scale=0.25, angle = -90]{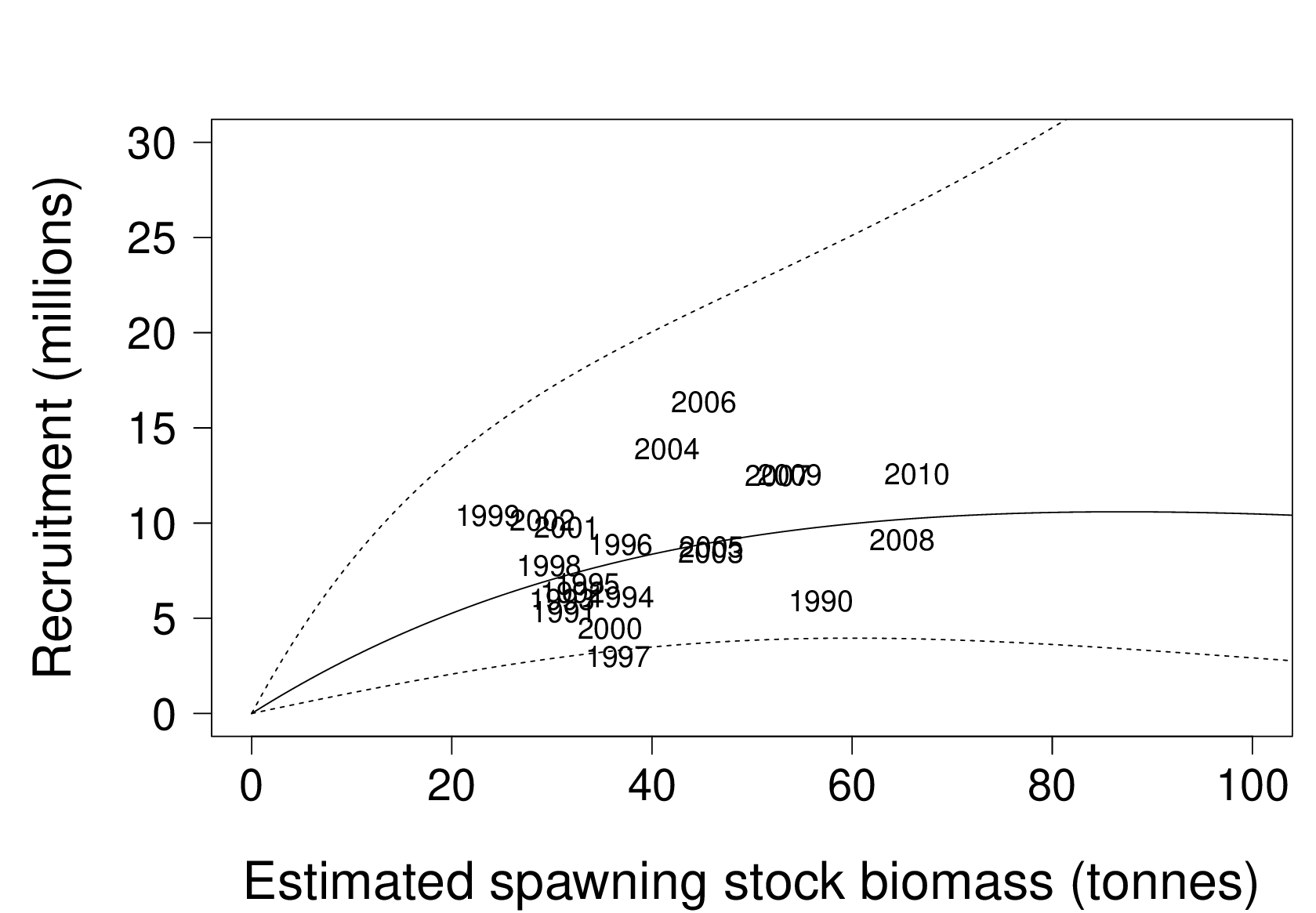}
       \end{minipage}
     }
\caption{}
       \label{fig:ModelOutput}
  \end{figure}

Catch of brown tiger prawns was characterised by a seasonal variation from a minimum in June--August to a maximum in January--March (Fig.~\ref{fig:ModelAndData}). The delay difference model captured both the intra-annual and inter-annual variability but systematically under-estimated large catches . Residuals were normally distributed with a slight tendency to increase over time and an intra-annual pattern was evident (Fig.~\ref{fig:ModelOutput}). Both discrepencies were small and did not influence the ranking of hypotheses presented below. Allowing catchability to vary seasonally as a function of temperature provided the largest improvement to the delay difference model (model 5 to 6 and model 3 to 4 in Tab.~\ref{tab:ModelComparison}). The second biggest improvement was achieved by differentiating catchability associated to targeted effort against non-targeted fishing (model 1 to 2 and model 6 to 4 ).  Relative fishing power corrections to all effort time series improved the fit to a lesser extent (model 2 to 3 and model 1 to 5) and reduced estimates of targeted catchability ($q_{1}$) by 12--17\%. The asymetric correction of targeted and non-targeted effort time series improved the fit slightly more (model 4 to 7).\\

The best description of tiger prawn catch was achieved by combining environmental and fishing effects into model 7. This model estimated targeted catchability equal to $q_{1} = 3.92 \pm 0.40 \ 10^{-4}$ boat-days$^{-1}$. A unit of non-targeted fishing effort was estimated to inflict around half the fishing mortality of a unit of targeted effort ($q_{2} = 1.91 \pm 0.24 \ 10^{-4}$ boat-days$^{-1}$). 90\% of recruitment to the fishery was estimated to occur between mid-November to the end of April and peak at the beginning of February (week 32, Fig.~\ref{fig:SeasonalDistOfRec}). Magnitude of recruitment increased, in average, by a factor of 1.8 before and after 2001 (Fig.~\ref{fig:TSrecruitment}). A linear regression between recruitment and spawning stock biomass (SSB) was not significant (P = 0.10). A fit of the Ricker stock-recruitment relationship showed that this aspect of the dynamic of the stock was the most uncertain (Fig.~\ref{fig:Ricker:a}).\\

\begin{figure}[h!]
     \subfigure[]{ 
       \label{fig:projections:a}
       \begin{minipage}[b]{0.95\textwidth}
       \includegraphics[scale=0.5, angle = -90]{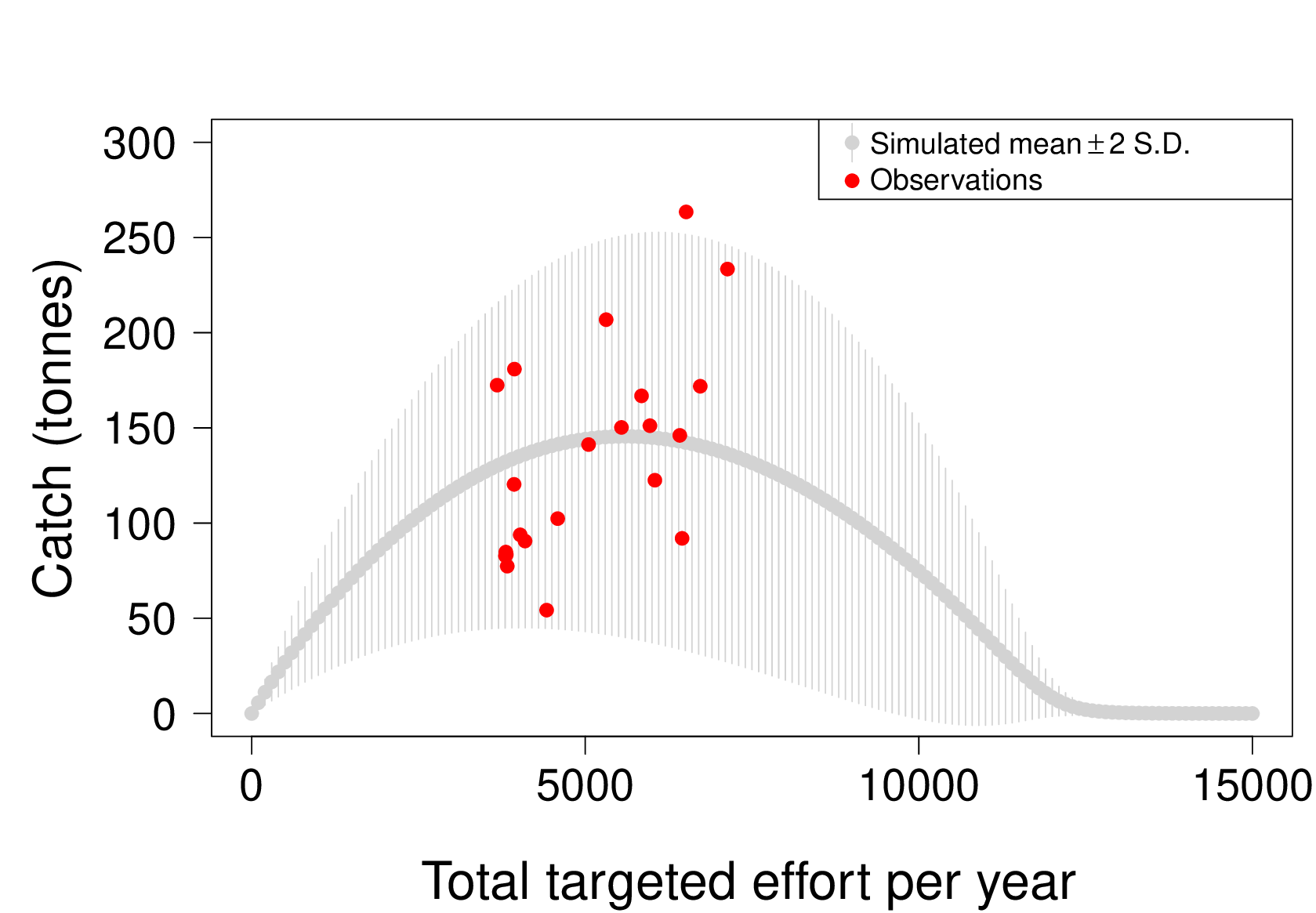}
       \end{minipage}
     }
     \subfigure[]{ 
       \label{fig:OverfishedOverfishing}
       \begin{minipage}[b]{0.95\textwidth}
       \includegraphics[scale=0.5, angle = -90]{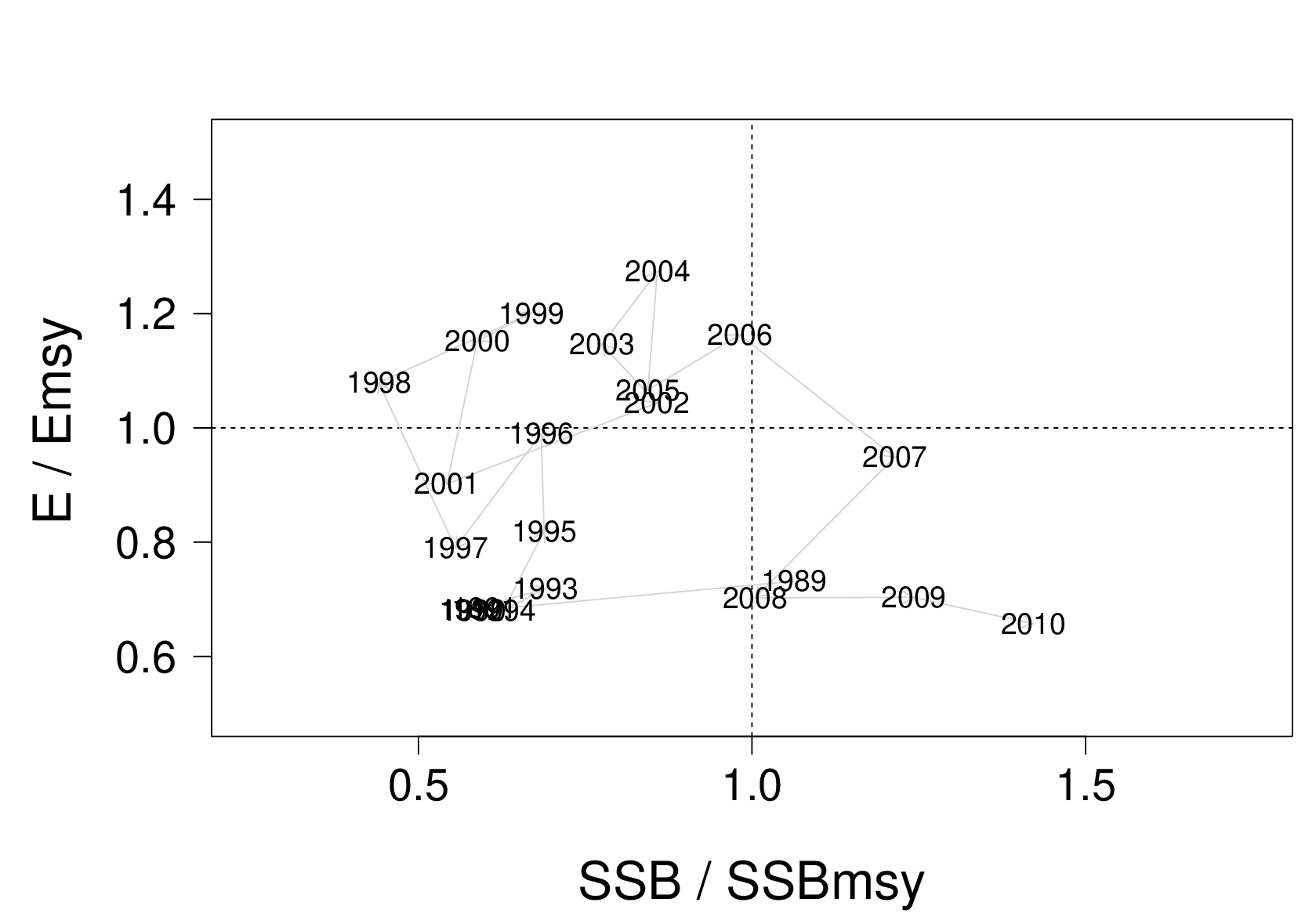}
       \end{minipage}
     }
\caption{}
       \label{fig:projections}
  \end{figure}

Projections of this stock model indicated that a maximum sustainable yield of 153 $\pm$ 50 tonnes can be achieved by applying 5600 targeted-fishing boat-days (in 1989 units), equivalent to 4000 boat-days in 2010 units (Fig.~\ref{fig:projections:a}). During most its recorded history, the stock of brown tiger prawn in Moreton has been overfished: its spawning stock biomass stayed around 60\% of SSB$_{{\rm MSY}}$ until 2001 and and then increased beyond this reference point (Fig.~\ref{fig:OverfishedOverfishing}). Data from 1989 were not alike those from the same period. Fishing effort was well below E$_{{\rm MSY}}$ at the beginning of the time series but increased up to 2000 without influencing much the SSB. Between 2000 and 2006, both SSB and fishing effort increased. After 2006, fishing mortality decreased while SSB increased to levels that were below (respectively above) those required to maintain maximum brown tiger prawn production in Moreton Bay. In 2010, the stock was not overfished nor was overfishing occuring.\\

\section{Discussion}

The delay difference model was implemented to quantify the impact of fishing on survival of brown tiger prawn in Moreton Bay. Defining fishing effort targeted at brown tiger prawn provided an important improvement to the fit, consistent with \cite{Zhoudoi:10.1139/f2011-052} recommending to weight these time-series differently in stock assessments. This model of the fishery can address the effect of shifting effort on and off tiger prawn but further research is needed to address shifts in effort between species in this multispecies fishery. Maximum likelihood estimate of targeted catchability ($3.92 \ 10^{-4}$ boat-days$^{-1}$) was about 5 times larger than 8.1--8.8 10$^{-5}$ estimated by \cite{Zhoudoi:10.1139/f2011-052} and \cite{Wang99a} in the NPF. This difference reflects to some extend the difference in stock size between these areas, Moreton Bay being 200 times smaller than the NPF. The model for stock-recruitment relationship was by far the most uncertain aspect of this fishery. The lack of significant linear relationship between recruitment and SSB is not un-expected given (a) the small number of observations and (b) the large variability in recruitment. The projections clearly encompassed the range of catch observed in this fishery over the past 20 years. \\

This stock assessment indicated that recent levels of exploitation of brown tiger prawn in Moreton Bay were sustainable. Increases in brown tiger prawn catch and catch rates were associated with declining fleet size and effort. The delay difference model estimated that recruitment increased simultaneously suggesting that brown tiger prawn has recovered from recruitment overfishing. Economics is the most likely driver of effort decline in this fishery: Australia almost doubled its import of prawn in the past 10 years, which account today for over 60\% of the total consumption of prawn in this country \citep{ABARE2011}. Imports of larger volumes of aquaculture production, in particular the white leg shrimp ({\it Penaeus vannamei}), have increased consumption and commercialization of species that were once primarily caught by local fishermen. The rapid decline of crustacean prices, 30--40\% \citep{Econsearch2010, Adams2005b, CurtottiEtAl2011Conf}, was strongly correlated with total effort in the fishery ($\rho = 0.79$). This accentuated competition in the seafood market affected the revenue of many fishing operations whose profits had already been eroded by years of increasing fuel prices \citep{ACS-ABARES2012, Sterling05r}. These two economic factors have certainly affected the fishery bionomic equilibrium \citep{clark1990mathematical} especially because trawling is an energy intensive fishing method. The changes observed in Moreton Bay trawl fishery in recent years are probably the result of aquaculture reducing pressure on wild fish stocks. The likelihood of fishing effort becoming a threat to the sustainability of brown tiger prawn harvest in Moreton Bay is low given that the fundamental economic drivers of the fishery are not likely to improve in a forseable future.\\

The simultaneous increase in SSB and effort targeted at tiger prawn between 2000 and 2006 was peculiar and suggested that factors other than those accounted for in this analysis might have affected the dynamic of the stock. Between those, a reduction of growth overfishing could explain a reduction in mortality and increase in brown tiger catch. The relative importance of this hypothesis was difficult to assess given the lack of information regarding the variation of catch's size-composition throughout the entire time-series. On the other hand, improvements in habitat, such as increased seagrass area or recent Marine Park closures in Moreton Bay, are unlikely to have contributed to the increase in {\it P. esculentus} population size. \cite{HylandR89} mapped the seagrass densities and distributions in Moreton Bay in the late 1980s and to our knowledge there are no evidence to suggest that these habitats have extended significantly since. In fact, large reclaimations of intertidal areas associated with expansion of the Port of Brisbane have probably contributed to a decline in such habitats for tiger prawns. The Queensland Government closed some area in the Bay to trawling in 2009 under the Moreton Bay Marine Park Plan.  However, while these areas may have had a positive impact on the brown tiger population size, they occurred after the dramatic decline in effort and after the population showed signs of recovery.\\

The estimated average rate of increase in fishing power in Moreton Bay of 1.7\% yr$^{-1}$ is a the higher end of the range estimated for the fleet operating on the east coast of Australia (0.5 -- 1.8 \% yr$^{-1}$) for the same period \citep{ONeill03a, ONeil2007a}) and at the lower end of envelop of the lower possible cases (1.8\% and 2.8\%) estimated by \citep{bis08a,Zhoudoi:10.1139/f2011-052} for 1980--2007. This difference might result from (a) an intrinsic difference in fishing improvements between smaller vessels ($<$ 14 m) operating in the sheltered waters of Moreton Bay compared to vessels operating in the Gulf of Carpentaria capable of long fishing trip requiring to withstand bad weather at sea; or (b) statistical treatment to allow for the possibility of greater impacts of technologies, referred to as the "high" treatment \citep{bis08a}. 
The steep increase in fishing power starting in 1995 corresponds to the adoption of GPS in the fleet \cite{ONeill03a}. The increase in fishing power starting in 2003 is not clearly associated with any technological improvement and might be the result of less efficient boat leaving the fishery.
The year term in catch rates standardizations \citep{maun04a} corresponds to an indice of abundance, providing the size of areas fished, and the spatial pattern of effort, have remained constant over the years \citep{bis08a}.  The present analysis did not account for spatial information which might result in interpreting wrongly variations of catch rates as a result of variations of abundance and fishing gear rather than a change in fishing locations \citep{cam04a, Wal03a}.\\

A range of hypotheses were compared to determine which influenced most the dynamic of this fishery. Temperature was found to determine the magnitude of catch by changing the duration of emergence of brown tiger prawns \citep{hil85a}. This behavioural change is related to the frequency of feeding that dependent on metabolic and digestive rates regulated by ambient temperatures in aquatic poikilotherms \citep{Fonds1992127}. The time-series of temperatures used in this analysis under-estimates the amplitude of variations in the Bay because shallow waters warm and cool faster in response air temperature than larger bodies of seawater. Such difference could explain the discrepancies between the model and the data evidenced by the residues's weekly pattern. Future research should include time-series of temperature collected {\it in-situ} and assess how they improve the fit of the delay difference model. Moreover this work provides a framework to evaluate the effect of climate change on the dynamic of tiger prawn fisheries. Given the present results, a rise in water temperature is expected to benefit fishermen in Moreton Bay by increasing brown tiger prawn catchability. The importance of this effect relative to, for example, fishing power increases could be quantified. Nevertheless, other effects of temperature such as its influence on timing and length of spawning season will need to be included to provide a comprehensive model of the effects of climate change on this tiger prawn fishery.\\

\section*{Acknowledgements}

This work was developed as part of the "Harvest strategy evaluations and co-management for the Moreton Bay Trawl Fishery" Cooperative Research Center project 2009/774. We are grateful for the many discussions with industry participants and representatives, in particular Dr. D. Sterling for estimating trawl swept area rates using his prawn trawl performance model and Mr M. Wood for promoting a better understanding of this fishery. We are grateful to Dr. G. Leigh, Dr. A. Campbell, Prof. Y.G. Wang and Dr. J. Robins for the many discussions on this topic. We thank Dr N. Wade, Dr. A. Norman and Dr W. N. Venables from the CSIRO for pointing relevant literature.

\clearpage
\newpage

\section*{Figures caption}
Fig.~\ref{fig:Map}: Top map: the spatial distribution of the brown tiger prawn ({\it Penaeus esculentus}) in Australia (from \cite{Grey83r}) with the location of Moreton Bay indicated by a black square. Bottom map: the location of trawling ground covering an area of about 800 km$^{2}$ (striped area). The dots represent areas closed to trawling including habitat protection zones, conservation and marine national parks.\\

Fig.~\ref{fig:TotalPrawnInfo}: Left: Recent trends in catch of all species of prawn and number of vessels fishing in Moreton Bay. Right: Species catch composition by year.\\

Fig.~\ref{fig:EstimatesOfAvailability}: Some relevant biological information: (a) proportion of mature female biomass; (b) duration of emergence determined experimentally by \cite{hil85a}; (c) monthly average seawater temperature in Moreton Bay and (d) estimated indices of availability of brown tiger prawn in Moreton Bay. \\

Fig.~\ref{fig:TargetvsNonTarget}: Relationship between total brown tiger prawn catch and effort (on the log-scale) by boat and year for fishing events targeted at brown tiger prawn or not.\\

Fig.~\ref{fig:TSCatchEffortAndCPUE}: Time series of brown tiger prawn ({\it Penaeus esculentus}) catch and catch per unit of effort in Moreton Bay (left panel); time series of effort catching tiger prawn in Moreton Bay (right panel).\\

Fig.~\ref{fig:FishPower}: (a) Estimated changes in fishing power relative to 1988 based on a GLM of brown tiger prawn catch. The vertical bars indicate 2 standard errors from the mean. (b) Variations of fishing effort corrected by fishing power.\\


Fig.~\ref{fig:ModelOutput}: (a) Time series of residuals of the delay difference model fit (b) Proportion of recruitment in each month estimated using the von Mises distribution. (c) Time series of recruitment estimated from model 3. (d) Estimated relationship between spawning stock biomass and recruitment fitted with a Ricker function. The dotted lines represent 95\% confidence interval of predictions from this model.\\

Fig.~\ref{fig:projections}: (a) Simulated long-term yield of brown tiger prawn at fixed level of effort to determine maximum sustainable yield (MSY). (b) Trajectory of the fishery through time in relation to spawning stock biomass at MSY (x-axis) and effort at MSY (y-axis). Note that effort (E) on this graph correspond to targeted effort corrected by both fishing power and availability.\\

\clearpage
\newpage

\section*{Figures}

\clearpage
\newpage
\section*{Tables}

\begin{table}[ht]
\begin{center}
\begin{tabular}{lll}
  \hline
Parameter & Value & Reference \\
  \hline
  $\rho$  & 0.963      & \cite{grib94a}\\
$w_{k-1}$  & 17.8 grams  & based on the method of \cite{sch85a} \\
$w_{k}$    & 19.5 grams  & based on the method of \cite{sch85a} \\
$M$       & 0.045 week$^{-1}$ & \cite{dichmont2003application}\\
   \hline
\end{tabular}
\caption{Values of parameters fixed in the delay difference model.}
\label{tab:ModelPar}
\end{center}
\end{table}

\begin{table}[ht]
\begin{center}
\begin{tabular}{ll}
  \toprule
Model & Fishing mortality \\
\hline
1 & $F_{t} = q_{1} \ E_{t}$ \\
2 & $F_{t} = q_{1} \ E_{t}(\rm{targeted}) + q_{2} \ E_{t}(\rm{non \ targeted})$ \\
3 & $F_{t} = \beta \ ( q_{1} \ E_{t}(\rm{targeted}) + q_{2} \ E_{t}(\rm{non \ targeted}) )$ \\
4 & $F_{t} = \gamma \ \beta \ ( q_{1} \ E_{t}(\rm{targeted}) + q_{2} \ E_{t}(\rm{non \ targeted}) )$\\
5 & $F_{t} = \beta \ q_{1} \ E_{t}$\\
6 & $F_{t} = \gamma \ \beta \ q_{1} \ E_{t}$\\
7 & $F_{t} = \gamma \ \beta \ q_{1} \ E_{t}(\rm{targeted}) + \gamma \ q_{2} \ E_{t}(\rm{non \ targeted}) $\\
  \bottomrule
\end{tabular}
\caption{Fishing mortality equations used in each delay difference model.}
\label{Tab:FishingMortalityModels}
\end{center}
\end{table}

\begin{table}[ht]
\centering
\begin{tabular}{lc}
  \hline
Definition of targeting & Res. SSQ \\ 
  \hline
tiger / total $>$ 0.1 & 18470 \\ 
tiger / total $>$ 0.2 & {\bf 17203} \\ 
tiger / total $>$ 0.3 & 17281 \\ 
tiger / total $>$ 0.4 & 17721 \\ 
tiger / total $>$ 0.5 & 18416 \\ 
tiger / total $>$ 0.6 & 19248 \\ 
tiger / total $>$ 0.7 & 20021 \\ 
tiger / total $>$ 0.8 & 21335 \\ 
tiger / total $>$ 0.9 & 22644 \\ 
tiger $>$ banana \& tiger $>$ greasyback \& tiger $>$ king & 18029 \\ 
banana = 0 \& king = 0 & 22583 \\ 
   \hline
\end{tabular}
\caption{Comparison of residual sum of squares (SSQ) of ANCOVAs of brown tiger prawn catch and effort on the log-scale for a variety of targeting definitions. The term "total" refers to the sum of tiger, banana, greasyback and eastern king prawns catch.} 
\label{tab:ANCOVA}
\end{table}

\begin{sidewaystable}[ht]
\centering
{\tiny
\begin{tabular}{llcccccccccccccc}
  \hline
 & \bigcell{c}{Boat \\ Mark} & \bigcell{c}{record \\ number} & \bigcell{c}{Swept \\ area} & \bigcell{c}{Lunar \\ Quarters} & \bigcell{c}{colour \\ echo--sounder} & satnav & GPS & dGPS & plotter & autopilot & \bigcell{c}{GPS coupled \\ autopilot} & \bigcell{c}{GPS coupled \\ radar} & \bigcell{c}{computer \\mapping} & BRD & TED \\ 
  \hline
  Boat Mark & 1.00 &  &  &  &  &  &  &  &  &  &  &  &  &  &  \\ 
  record number &  & 1.00 &  &  &  &  &  &  &  &  &  &  &  &  &  \\ 
  Swept area & 0.34 & 0.35 & 1.00 & 0.00 & 0.01 & 0.04 & 0.02 & 0.00 & 0.02 & 0.02 & 0.09 & 0.00 & 0.02 & 0.00 & 0.00 \\ 
  Lunar Quarters &  &  &  & 1.00 &  &  &  &  &  &  &  &  &  &  &  \\ 
  colour echo--sounder & 0.95 & 0.95 & 0.01 & 0.00 & 1.00 & 0.01 & 0.00 & 0.01 & 0.01 & 0.42 & 0.00 & 0.01 & 0.10 & 0.01 & 0.00 \\ 
  satnav & 0.74 & 1.00 & 0.04 & 0.00 & 0.01 & 1.00 & 0.00 & 0.01 & 0.01 & 0.00 & 0.11 & 0.00 & 0.07 & 0.03 & 0.00 \\ 
  GPS & 0.69 & 0.79 & 0.02 & 0.00 & 0.00 & 0.00 & 1.00 & 0.00 & 0.19 & 0.00 & 0.05 & 0.01 & 0.10 & 0.04 & 0.00 \\ 
  dGPS & 0.89 & 0.93 & 0.00 & 0.00 & 0.01 & 0.01 & 0.00 & 1.00 & 0.02 & 0.01 & 0.00 & 0.00 & 0.06 & 0.01 & 0.00 \\ 
  plotter & 0.82 & 0.92 & 0.02 & 0.00 & 0.01 & 0.01 & 0.19 & 0.02 & 1.00 & 0.03 & 0.00 & 0.01 & 0.00 & 0.00 & 0.00 \\ 
  autopilot & 0.95 & 0.95 & 0.02 & 0.00 & 0.42 & 0.00 & 0.00 & 0.01 & 0.03 & 1.00 & 0.03 & 0.01 & 0.05 & 0.00 & 0.00 \\ 
  GPS coupled autopilot & 0.93 & 0.93 & 0.09 & 0.00 & 0.00 & 0.11 & 0.05 & 0.00 & 0.00 & 0.03 & 1.00 & 0.09 & 0.18 & 0.01 & 0.01 \\ 
  GPS coupled radar & 0.62 & 0.62 & 0.00 & 0.00 & 0.01 & 0.00 & 0.01 & 0.00 & 0.01 & 0.01 & 0.09 & 1.00 & 0.02 & 0.00 & 0.00 \\ 
  computer.mapping & 0.83 & 0.93 & 0.02 & 0.00 & 0.10 & 0.07 & 0.10 & 0.06 & 0.00 & 0.05 & 0.18 & 0.02 & 1.00 & 0.07 & 0.02 \\ 
  BRD & 0.47 & 0.55 & 0.00 & 0.00 & 0.01 & 0.03 & 0.04 & 0.01 & 0.00 & 0.00 & 0.01 & 0.00 & 0.07 & 1.00 & 0.02 \\ 
  TED & 0.40 & 0.54 & 0.00 & 0.00 & 0.00 & 0.00 & 0.00 & 0.00 & 0.00 & 0.00 & 0.01 & 0.00 & 0.02 & 0.02 & 1.00 \\ 
   \hline
\end{tabular}
}
\caption{Proportion of variability explained ($R^{2}$) by a pair-wise linear regression using a single variable (in rows) as the dependent variable and a single variable (in column) as the explanatory variable.} 
\label{tab:MultiCollinearity-LMonPairsOfVariable}
\end{sidewaystable}

\begin{table}[ht]
\centering
\begin{tabular}{lcccc}
  \hline
 & Df & Deviance & scaled dev. & Pr($>$ $\chi^{2}$) \\ 
  \hline
$<$none$>$ &  & 5458.53 &  &  \\ 
  log(SA) & 1 & 5908.35 & 1170.22 & 0.0000 \\ 
  BRD & 1 & 5462.25 & 9.69 & 0.0019 \\ 
  TED & 1 & 5460.71 & 5.69 & 0.0171 \\ 
  colour echo sounder & 1 & 5461.75 & 8.38 & 0.0038 \\ 
  dGPS & 1 & 5461.95 & 8.91 & 0.0028 \\ 
  satnav & 1 & 5499.25 & 105.94 & 0.0000 \\ 
  plotter & 1 & 5482.69 & 62.87 & 0.0000 \\ 
  GPSCoupledautopilot & 1 & 5460.90 & 6.18 & 0.0129 \\ 
  computer mapping & 1 & 5507.21 & 126.65 & 0.0000 \\ 
  lunar & 1 & 5462.11 & 9.32 & 0.0023 \\ 
  Year:Month & 244 & 10562.27 & 13277.33 & 0.0000 \\ 
   \hline
\end{tabular}
\caption{Assessment of the effect of removing a single variable from the full GLM}
\label{tab:Drop1GLM}
\end{table}

\begin{table}[ht]
\centering
\begin{tabular}{lcccc}
  \hline
 & Estimate & Std. Error & t value & Pr($>$$|$t$|$) \\ 
  \hline
log(SA) & 0.6384 & 0.0175 & 36.55 & 0.0000 \\ 
  BRD & 0.0955 & 0.0308 & 3.10 & 0.0019 \\ 
  TED & -0.0907 & 0.0378 & -2.40 & 0.0165 \\ 
  colour echo sounder & 0.0579 & 0.0193 & 3.00 & 0.0027 \\ 
  dGPS & 0.0688 & 0.0230 & 2.99 & 0.0028 \\ 
  satnav & 0.2254 & 0.0217 & 10.37 & 0.0000 \\ 
  plotter & 0.1104 & 0.0136 & 8.13 & 0.0000 \\ 
  GPSCoupledautopilot & -0.0332 & 0.0130 & -2.56 & 0.0106 \\ 
  computer mapping & 0.1521 & 0.0133 & 11.41 & 0.0000 \\ 
  lunar & -0.0450 & 0.0147 & -3.07 & 0.0021 \\ 
  Year1990:Month01 & -0.8020 & 0.6268 & -1.28 & 0.2008 \\ 
  $\ldots$         & $\ldots$ & $\ldots$ & $\ldots$ & $\ldots$ \\
  \hline
\end{tabular}
\caption{Parameter estimates for the GLM co-variates.}
\label{tab:FishPowerParEst}
\end{table}

\begin{sidewaystable}[ht]
\centering
\begin{tabular}{lcccccccc}
  \toprule
  Model & $-\log(L)$ & \multicolumn{2}{c}{Catchability} & \multicolumn{2}{c}{Recruitment distribution} & \multicolumn{2}{c}{Biomass} &  \multicolumn{1}{c}{}\\  \cmidrule(r){1-1}
  \cmidrule(r){2-2}
  \cmidrule(r){3-4}
  \cmidrule(r){5-6}
  \cmidrule(r){7-8}
  \cmidrule(r){9-9}
  & & $q_{1} \ (\times 10^{-4})$ & $q_{2} \ (\times 10^{-4})$ & $\mu$ & $\kappa$ & B(1) & B(2) & $\sigma$\\
  \midrule
   7 & 3659.05 & 3.92 $\pm$ 0.4 & 1.91 $\pm$ 0.24 & 0.64 $\pm$ 0.03 & 2.02 $\pm$ 0.08 & 0.22 $\pm$ 0.01 & 0.3 $\pm$ 0.01 & 5.93 $\pm$ 0.13 \\ 
    4 & 3669.41 & 4.02 $\pm$ 0.42 & 1.7 $\pm$ 0.22 & 0.65 $\pm$ 0.03 & 2.04 $\pm$ 0.09 & 0.22 $\pm$ 0.02 & 0.3 $\pm$ 0.02 & 5.98 $\pm$ 0.13 \\ 
    6 & 3707.42 & 1.51 $\pm$ 0.19 &  & 0.51 $\pm$ 0.02 & 5.32 $\pm$ 0.58 & 0.5 $\pm$ 0.29 & 0.63 $\pm$ 0.26 & 6.18 $\pm$ 0.13 \\ 
    3 & 3759.51 & 3.96 $\pm$ 0.31 & 1.76 $\pm$ 0.2 & 0.49 $\pm$ 0.03 & 2.38 $\pm$ 0.12 & 0.06 $\pm$ 0.01 & 0.13 $\pm$ 0.01 & 6.47 $\pm$ 0.14 \\ 
    2 & 3764.42 & 4.9 $\pm$ 0.38 & 1.87 $\pm$ 0.22 & 0.46 $\pm$ 0.03 & 2.27 $\pm$ 0.12 & 0.05 $\pm$ 0.05 & 0.12 $\pm$ 0.05 & 6.5 $\pm$ 0.14 \\ 
    5 & 3801.99 & 3 $\pm$ 0.27 &  & 0.5 $\pm$ 0.02 & 3.46 $\pm$ 0.27 & 0.06 $\pm$ 0.06 & 0.13 $\pm$ 0.05 & 6.72 $\pm$ 0.14 \\ 
    1 & 3819.91 & 3.38 $\pm$ 0.27 &  & 0.48 $\pm$ 0.02 & 3.65 $\pm$ 0.29 & 0.06 $\pm$ 0.06 & 0.12 $\pm$ 0.05 & 6.82 $\pm$ 0.15 \\ 
   \hline
\end{tabular}
\caption{Comparison of the negative log-likelihood and parameters estimates of different models. The results are ordered by increasing value of negative log-likelihood from top to bottom.} 
\label{tab:ModelComparison}
\end{sidewaystable}


\begin{thebibliography}{51}
\expandafter\ifx\csname natexlab\endcsname\relax\def\natexlab#1{#1}\fi
\providecommand{\url}[1]{\texttt{#1}}
\providecommand{\href}[2]{#2}
\providecommand{\path}[1]{#1}
\providecommand{\DOIprefix}{doi:}
\providecommand{\ArXivprefix}{arXiv:}
\providecommand{\URLprefix}{URL: }
\providecommand{\Pubmedprefix}{pmid:}
\providecommand{\doi}[1]{\href{http://dx.doi.org/#1}{\path{#1}}}
\providecommand{\Pubmed}[1]{\href{pmid:#1}{\path{#1}}}
\providecommand{\bibinfo}[2]{#2}
\ifx\xfnm\relax \def\xfnm[#1]{\unskip,\space#1}\fi
\bibitem[{Adams et~al.(2005)Adams, Keithly and Versaggi}]{Adams2005b}
\bibinfo{author}{Adams, C.}, \bibinfo{author}{Keithly, W.},
  \bibinfo{author}{Versaggi, S.}, \bibinfo{year}{2005}.
\newblock \bibinfo{title}{The shrimp import controversy}, in:
  \bibinfo{editor}{Schmitz, A.}, \bibinfo{editor}{Moss, C.B.},
  \bibinfo{editor}{Schmitz, T.G.}, \bibinfo{editor}{Koo, W.W.} (Eds.),
  \bibinfo{booktitle}{International Agriculture Trade Disputes: Case Studies In
  North America}. \bibinfo{publisher}{Calgary: University of Calgary Press}.
  chapter~\bibinfo{chapter}{15}, pp. \bibinfo{pages}{225--244}.
\bibitem[{Anon.(2010)}]{Econsearch2010}
\bibinfo{author}{Anon.}, \bibinfo{year}{2010}.
\newblock \bibinfo{title}{Economic Indicators for the Spencer Gulf and West
  Coast Prawn Fisheries}.
\newblock \bibinfo{type}{Technical Report}. Econsearch.
\bibitem[{Anon.(2011)}]{ABARE2011}
\bibinfo{author}{Anon.}, \bibinfo{year}{2011}.
\newblock \bibinfo{title}{Australian fisheries statistics 2011}.
\newblock Dept. of Agriculture, Fisheries and Forestry,
  \bibinfo{publisher}{Australian Government}.
\bibitem[{Anon.(2012a)}]{ACS-ABARES2012}
\bibinfo{author}{Anon.}, \bibinfo{year}{2012}a.
\newblock \bibinfo{title}{Agricultural commodity statistics 2012}.
\newblock Dept. of Agriculture, Fisheries and Forestry,
  \bibinfo{publisher}{Australian Government}.
\bibitem[{Anon.(2012b)}]{metoc2012}
\bibinfo{author}{Anon.}, \bibinfo{year}{2012}b.
\newblock \bibinfo{title}{Hydrography, Meteorology and Oceanography services}.
\newblock \bibinfo{organization}{Australian Government -- Department of
  Defence}.
\newblock \URLprefix \url{www.metoc.gov.au}. \bibinfo{note}{last accessed on
  9$^{th}$ January 2012}.
\bibitem[{Bevington and Robinson(2003)}]{bevrob03}
\bibinfo{author}{Bevington, P.R.}, \bibinfo{author}{Robinson, D.K.},
  \bibinfo{year}{2003}.
\newblock \bibinfo{title}{Data reduction and error analysis}.
\newblock \bibinfo{edition}{3rd} ed., \bibinfo{publisher}{Mac Graw Hill}.
\bibitem[{Bishop et~al.(2008)Bishop, Venables, Dichmont and J.}]{bis08a}
\bibinfo{author}{Bishop, J.}, \bibinfo{author}{Venables, W.N.},
  \bibinfo{author}{Dichmont, C.M.}, \bibinfo{author}{J., S.D.},
  \bibinfo{year}{2008}.
\newblock \bibinfo{title}{Standardizing catch rates: is logbook information by
  itself enough ?}
\newblock \bibinfo{journal}{ICES Journal of Marine Science}
  \bibinfo{volume}{65}, \bibinfo{pages}{255--266}.
\bibitem[{Brun and Rademakers(1997)}]{root}
\bibinfo{author}{Brun, R.}, \bibinfo{author}{Rademakers, F.},
  \bibinfo{year}{1997}.
\newblock \bibinfo{title}{Root - an object oriented data analysis framework},
  in: \bibinfo{booktitle}{Proceedings AIHENP'96 Workshop, Lausanne, Sep. 1996,
  Nucl. Inst. \& Meth. in Phys. Res. A 389}, pp. \bibinfo{pages}{81--86}.
\newblock \bibinfo{note}{See also http://root.cern.ch/}.
\bibitem[{Burnham and Anderson(2003)}]{Burnb03}
\bibinfo{author}{Burnham, K.P.}, \bibinfo{author}{Anderson, D.},
  \bibinfo{year}{2003}.
\newblock \bibinfo{title}{Model Selection and Multi-Model Inference}.
\newblock \bibinfo{edition}{2nd} ed.,
  \bibinfo{publisher}{Spring{\-}er-Ver{\-}lag}.
\bibitem[{Campbell(2004)}]{cam04a}
\bibinfo{author}{Campbell, R.A.}, \bibinfo{year}{2004}.
\newblock \bibinfo{title}{\uppercase{cpue} standardisation and the construction
  of indices of stock abundance in a spatially varying fishery using general
  linear models}.
\newblock \bibinfo{journal}{Fisheries Research} \bibinfo{volume}{70},
  \bibinfo{pages}{2009--227}.
\bibitem[{Clark(1990)}]{clark1990mathematical}
\bibinfo{author}{Clark, C.}, \bibinfo{year}{1990}.
\newblock \bibinfo{title}{Mathematical bioeconomics: the optimal management of
  renewable resources}.
\newblock Pure and applied mathematics, \bibinfo{publisher}{Wiley}.
\bibitem[{Courtney et~al.(2012)Courtney, Kienzle, Pascoe, O'Neill, Leigh, Wang,
  Innes, Landers, Braccini, Prosser, Baxter, Sterling and Larkin}]{SeaCRC2012}
\bibinfo{author}{Courtney, A.}, \bibinfo{author}{Kienzle, M.},
  \bibinfo{author}{Pascoe, S.}, \bibinfo{author}{O'Neill, M.},
  \bibinfo{author}{Leigh, G.}, \bibinfo{author}{Wang, Y.G.},
  \bibinfo{author}{Innes, J.}, \bibinfo{author}{Landers, M.},
  \bibinfo{author}{Braccini, M.}, \bibinfo{author}{Prosser, A.},
  \bibinfo{author}{Baxter, P.}, \bibinfo{author}{Sterling, D.},
  \bibinfo{author}{Larkin, J.}, \bibinfo{year}{2012}.
\newblock \bibinfo{title}{Harvest strategy evaluations and co-management for
  the Moreton Bay Trawl Fishery}.
\newblock \bibinfo{type}{Technical Report} \bibinfo{number}{Project 2009/774}.
  Australian Seafood CRC.
\bibitem[{Courtney and Masel(1997)}]{court97a}
\bibinfo{author}{Courtney, A.}, \bibinfo{author}{Masel, J.},
  \bibinfo{year}{1997}.
\newblock \bibinfo{title}{Spawning stock dynamics of two penaeid prawns, {\it
  \uppercase{m}etapenaeus bennettae} and {\it \uppercase{p}enaeus esculentus},
  in \uppercase{M}oreton \uppercase{B}ay, \uppercase{Q}ueensland,
  \uppercase{A}ustralia}.
\newblock \bibinfo{journal}{Marine Ecology Progress Series}
  \bibinfo{volume}{148}, \bibinfo{pages}{37--47}.
\bibitem[{Courtney et~al.(1995)Courtney, Masel and Die}]{court95a}
\bibinfo{author}{Courtney, A.}, \bibinfo{author}{Masel, J.},
  \bibinfo{author}{Die, D.}, \bibinfo{year}{1995}.
\newblock \bibinfo{title}{Temporal and spatial patterns in recruitment of three
  penaeid prawns in \uppercase{m}oreton \uppercase{b}ay,
  \uppercase{q}ueensland, \uppercase{a}ustralia}.
\newblock \bibinfo{journal}{Estuarine, Coastal and Shelf Science}
  \bibinfo{volume}{41}.
\bibitem[{Curtotti et~al.(2011)Curtotti, Hormis, McGill, Pham, Vieira, Perks
  and George}]{CurtottiEtAl2011Conf}
\bibinfo{author}{Curtotti, R.}, \bibinfo{author}{Hormis, M.},
  \bibinfo{author}{McGill, K.}, \bibinfo{author}{Pham, T.},
  \bibinfo{author}{Vieira, S.}, \bibinfo{author}{Perks, C.},
  \bibinfo{author}{George, D.}, \bibinfo{year}{2011}.
\newblock \bibinfo{title}{Australian fisheries - outlook and economic
  indicators}, in: \bibinfo{booktitle}{Outlook conference},
  \bibinfo{organization}{Australian Bureau of Agricultural and Resource
  Economics and Sciences (ABARES)}.
\bibitem[{Davie et~al.(2011)Davie, Cranitch, Wright and
  Cowell}]{nla.cat-vn5267095}
\bibinfo{author}{Davie, P.}, \bibinfo{author}{Cranitch, G.},
  \bibinfo{author}{Wright, J.}, \bibinfo{author}{Cowell, B.},
  \bibinfo{year}{2011}.
\newblock \bibinfo{title}{Wild guide to Moreton Bay and adjacent coasts}.
\newblock \bibinfo{edition}{2nd} ed., \bibinfo{publisher}{Queensland Museum,
  Brisbane, Australia}.
\bibitem[{Deriso(1980)}]{Der80a}
\bibinfo{author}{Deriso, R.}, \bibinfo{year}{1980}.
\newblock \bibinfo{title}{Harvesting strategies and parameter estimation for an
  age-structured model}.
\newblock \bibinfo{journal}{Canadian Journal of Fisheries and Aquatic Sciences}
  \bibinfo{volume}{37}, \bibinfo{pages}{268--282}.
\bibitem[{Dichmont et~al.(2006)Dichmont, Deng, Punt, Venables and
  Haddon}]{Dichmont2006204}
\bibinfo{author}{Dichmont, C.}, \bibinfo{author}{Deng, A.},
  \bibinfo{author}{Punt, A.}, \bibinfo{author}{Venables, W.},
  \bibinfo{author}{Haddon, M.}, \bibinfo{year}{2006}.
\newblock \bibinfo{title}{Management strategies for short-lived species: The
  case of \uppercase{A}ustralia's \uppercase{N}orthern \uppercase{P}rawn
  \uppercase{F}ishery: 1. \uppercase{A}ccounting for multiple species, spatial
  structure and implementation uncertainty when evaluating risk}.
\newblock \bibinfo{journal}{Fisheries Research} \bibinfo{volume}{82},
  \bibinfo{pages}{204--220}.
\bibitem[{Dichmont et~al.(2003)Dichmont, Punt, Deng, Dell and
  Venables}]{dichmont2003application}
\bibinfo{author}{Dichmont, C.}, \bibinfo{author}{Punt, A.},
  \bibinfo{author}{Deng, A.}, \bibinfo{author}{Dell, Q.},
  \bibinfo{author}{Venables, W.}, \bibinfo{year}{2003}.
\newblock \bibinfo{title}{Application of a weekly delay-difference model to
  commercial catch and effort data for tiger prawns in australia’s northern
  prawn fishery}.
\newblock \bibinfo{journal}{Fisheries Research} \bibinfo{volume}{65},
  \bibinfo{pages}{335--350}.
\bibitem[{Draper and Smith(1998)}]{drapb}
\bibinfo{author}{Draper, N.}, \bibinfo{author}{Smith, H.},
  \bibinfo{year}{1998}.
\newblock \bibinfo{title}{Applied Regression Analysis}.
\newblock \bibinfo{edition}{3$^{rd}$} ed., \bibinfo{publisher}{Wiley \& Sons}.
\bibitem[{Fonds et~al.(1992)Fonds, Cronie, Vethaak and Puyl}]{Fonds1992127}
\bibinfo{author}{Fonds, M.}, \bibinfo{author}{Cronie, R.},
  \bibinfo{author}{Vethaak, A.}, \bibinfo{author}{Puyl, P.V.D.},
  \bibinfo{year}{1992}.
\newblock \bibinfo{title}{Metabolism, food consumption and growth of plaice
  (pleuronectes platessa) and flounder (platichthys flesus) in relation to fish
  size and temperature}.
\newblock \bibinfo{journal}{Netherlands Journal of Sea Research}
  \bibinfo{volume}{29}, \bibinfo{pages}{127--143}.
\bibitem[{Garc\'{i}a and Reste(1981)}]{garcia1981life}
\bibinfo{author}{Garc\'{i}a, S.}, \bibinfo{author}{Reste, L.},
  \bibinfo{year}{1981}.
\newblock \bibinfo{title}{Life cycles, dynamics, exploitation and management of
  coastal penaeid shrimp stocks}.
\newblock FAO fisheries technical paper, \bibinfo{publisher}{FAO}.
\bibitem[{Grey et~al.(1983)Grey, Dall and Baker}]{Grey83r}
\bibinfo{author}{Grey, D.}, \bibinfo{author}{Dall, W.}, \bibinfo{author}{Baker,
  A.}, \bibinfo{year}{1983}.
\newblock \bibinfo{title}{A guide to the Australian Penaeid Prawns}.
\newblock \bibinfo{type}{Technical Report} \bibinfo{number}{2005/239}.
  Department of Primary Production.
\bibitem[{Gribble and Dredge(1994)}]{grib94a}
\bibinfo{author}{Gribble, N.}, \bibinfo{author}{Dredge, M.},
  \bibinfo{year}{1994}.
\newblock \bibinfo{title}{Mixed-species yield-per-recruit simulations of the
  effect of seasonal closure on a central queensland coastal prawn trawling
  ground}.
\newblock \bibinfo{journal}{Canadian Journal of Fisheries and Aquatic Sciences}
  \bibinfo{volume}{51}, \bibinfo{pages}{998--1010}.
\bibitem[{Haddon(2010)}]{haddon2010modelling}
\bibinfo{author}{Haddon, M.}, \bibinfo{year}{2010}.
\newblock \bibinfo{title}{Modelling and quantitative methods in fisheries}.
\newblock \bibinfo{publisher}{Chapman \& Hall/CRC}.
\bibitem[{Hilborn and Walters(1992)}]{hil92b}
\bibinfo{author}{Hilborn, R.}, \bibinfo{author}{Walters, C.},
  \bibinfo{year}{1992}.
\newblock \bibinfo{title}{Quantitative fisheries stock assessment: choice,
  dynamics and uncertainty}.
\newblock \bibinfo{publisher}{Chapman and Hall}.
\bibitem[{Hill(1985)}]{hil85a}
\bibinfo{author}{Hill, B.}, \bibinfo{year}{1985}.
\newblock \bibinfo{title}{Effect of temperature on duration of emergence, speed
  of movement, and catchability of the prawn {\it \uppercase{p}enaeus
  esculentus}}, in: \bibinfo{editor}{Rothlisberg~PC, Hill~BJ, S.D.} (Ed.),
  \bibinfo{booktitle}{Proceedings of the Second Australian National Prawn
  Seminar}, \bibinfo{publisher}{Simpson Halligan, Brisbane, Australia}. pp.
  \bibinfo{pages}{77--83}.
\bibitem[{Hyland(1987)}]{Hyland87PhD}
\bibinfo{author}{Hyland, S.}, \bibinfo{year}{1987}.
\newblock \bibinfo{title}{An investigation of the nektobenthic organisms in
  Logan River and Moreton Bay (Queensland) with an emphasis on penaeid prawns}.
\newblock Ph.D. thesis. University of Queensland.
\bibitem[{Hyland et~al.(1989)Hyland, Courtney and Butler}]{HylandR89}
\bibinfo{author}{Hyland, S.J.}, \bibinfo{author}{Courtney, A.J.},
  \bibinfo{author}{Butler, C.T.}, \bibinfo{year}{1989}.
\newblock \bibinfo{title}{Distribution of seagrass in the Moreton Region from
  Coolangatta to Noosa}.
\newblock \bibinfo{type}{Information Series} \bibinfo{number}{QI89010}.
  Queensland Dept of Primary Industry.
\bibitem[{James and Winkler(2004)}]{minuit2}
\bibinfo{author}{James, F.}, \bibinfo{author}{Winkler, M.},
  \bibinfo{year}{2004}.
\newblock \bibinfo{title}{Minuit user's guide}.
\newblock \bibinfo{howpublished}{http://www.cern.ch/minuit}.
\bibitem[{Keys(2003)}]{Keys2003325}
\bibinfo{author}{Keys, S.}, \bibinfo{year}{2003}.
\newblock \bibinfo{title}{Aspects of the biology and ecology of the brown tiger
  prawn, penaeus esculentus, relevant to aquaculture}.
\newblock \bibinfo{journal}{Aquaculture} \bibinfo{volume}{217},
  \bibinfo{pages}{325 -- 334}.
\bibitem[{Mardia and Jupp(1999)}]{mardia1999directional}
\bibinfo{author}{Mardia, K.}, \bibinfo{author}{Jupp, P.}, \bibinfo{year}{1999}.
\newblock \bibinfo{title}{Directional statistics}.
\newblock \bibinfo{publisher}{Wiley}.
\bibitem[{Maunder and Punt(2004)}]{maun04a}
\bibinfo{author}{Maunder, M.N.}, \bibinfo{author}{Punt, A.E.},
  \bibinfo{year}{2004}.
\newblock \bibinfo{title}{Standardizing catch and effort data: a review of
  recent approaches}.
\newblock \bibinfo{journal}{Fisheries Research} \bibinfo{volume}{70},
  \bibinfo{pages}{141--179}.
\bibitem[{O'Brien(1994)}]{OBrien94a}
\bibinfo{author}{O'Brien, C.}, \bibinfo{year}{1994}.
\newblock \bibinfo{title}{Population dynamics of juvenile tiger prawns {\it
  \uppercase{p}enaeus esculentus} in south \uppercase{Q}ueensland,
  \uppercase{A}ustralia}.
\newblock \bibinfo{journal}{Marine Ecology Progress Series}
  \bibinfo{volume}{104}, \bibinfo{pages}{247--256}.
\bibitem[{O'Neill et~al.(2003)O'Neill, Courtney, Turnbull, Good, Yeomans,
  Staunton~Smith and Shootingstar}]{ONeill03a}
\bibinfo{author}{O'Neill, M.}, \bibinfo{author}{Courtney, A.},
  \bibinfo{author}{Turnbull, C.}, \bibinfo{author}{Good, N.},
  \bibinfo{author}{Yeomans, K.}, \bibinfo{author}{Staunton~Smith, J.},
  \bibinfo{author}{Shootingstar, C.}, \bibinfo{year}{2003}.
\newblock \bibinfo{title}{Comparison of relative fishing power between
  different sectors of the queensland trawl fishery, australia}.
\newblock \bibinfo{journal}{Fisheries Research} , \bibinfo{pages}{309--321}.
\bibitem[{O'Neill and Leigh(2007)}]{ONeil2007a}
\bibinfo{author}{O'Neill, M.}, \bibinfo{author}{Leigh, G.},
  \bibinfo{year}{2007}.
\newblock \bibinfo{title}{Fishing power increases continues in
  \uppercase{Q}ueensland's east coast trawl fishery}.
\newblock \bibinfo{journal}{Fisheries Research} \bibinfo{volume}{85},
  \bibinfo{pages}{84--92}.
\bibitem[{O'Neill and Turnbull(2006)}]{ONeill2006R}
\bibinfo{author}{O'Neill, M.}, \bibinfo{author}{Turnbull, C.},
  \bibinfo{year}{2006}.
\newblock \bibinfo{title}{Stock assessment of the Torres Strait tiger prawn
  fishery ({\it Penaeus esculentus})}.
\newblock \bibinfo{type}{Information Series} \bibinfo{number}{QI05132}.
  Queensland Dept of Primary Industry and Fisheries.
\bibitem[{Ovenden et~al.(2007)Ovenden, Peel, Street, Courtney, Hoyle, Peel and
  Podlich}]{MEC:MEC3132}
\bibinfo{author}{Ovenden, J.}, \bibinfo{author}{Peel, D.},
  \bibinfo{author}{Street, R.}, \bibinfo{author}{Courtney, A.},
  \bibinfo{author}{Hoyle, S.}, \bibinfo{author}{Peel, S.},
  \bibinfo{author}{Podlich, H.}, \bibinfo{year}{2007}.
\newblock \bibinfo{title}{The genetic effective and adult census size of an
  \uppercase{A}ustralian population of tiger prawns ({\it \uppercase{p}enaeus
  esculentus})}.
\newblock \bibinfo{journal}{Molecular Ecology} \bibinfo{volume}{16},
  \bibinfo{pages}{127--138}.
\bibitem[{Parke(2013)}]{parke2013b}
\bibinfo{author}{Parke, J.}, \bibinfo{year}{2013}.
\newblock \bibinfo{title}{Against the tide: \uppercase{Q}ueensland's
  \uppercase{M}oreton \uppercase{B}ay fishing industry since 1824}.
\newblock \bibinfo{publisher}{5Word Productions}.
\bibitem[{Pascoe et~al.(2013)Pascoe, Dichmont, Brooks, Pears and
  Jebreen}]{Pascoe2013115}
\bibinfo{author}{Pascoe, S.}, \bibinfo{author}{Dichmont, C.},
  \bibinfo{author}{Brooks, K.}, \bibinfo{author}{Pears, R.},
  \bibinfo{author}{Jebreen, E.}, \bibinfo{year}{2013}.
\newblock \bibinfo{title}{Management objectives of queensland fisheries:
  Putting the horse before the cart}.
\newblock \bibinfo{journal}{Marine Policy} \bibinfo{volume}{37},
  \bibinfo{pages}{115--122}.
\bibitem[{Quinn and Deriso(1999)}]{quin99b}
\bibinfo{author}{Quinn, T.J.}, \bibinfo{author}{Deriso, R.B.},
  \bibinfo{year}{1999}.
\newblock \bibinfo{title}{Quantitative fish dynamics}.
\newblock \bibinfo{publisher}{Oxford University Press}.
\bibitem[{{R Core Team}(2013)}]{R}
\bibinfo{author}{{R Core Team}}, \bibinfo{year}{2013}.
\newblock \bibinfo{title}{R: A Language and Environment for Statistical
  Computing}.
\newblock \bibinfo{organization}{R Foundation for Statistical Computing}.
  \bibinfo{address}{Vienna, Austria}.
\newblock \URLprefix \url{http://www.R-project.org/}.
\bibitem[{Robins et~al.(1998)Robins, Wang and Die}]{rob98a}
\bibinfo{author}{Robins, C.M.}, \bibinfo{author}{Wang, Y.},
  \bibinfo{author}{Die, D.}, \bibinfo{year}{1998}.
\newblock \bibinfo{title}{The impact of global positioning systems and plotters
  on fishing power in the \uppercase{N}orthern \uppercase{P}rawn
  \uppercase{F}ishery}.
\newblock \bibinfo{journal}{Canadian Journal of Fisheries and Aquatic Sciences}
  \bibinfo{volume}{55}.
\bibitem[{Schnute(1985)}]{sch85a}
\bibinfo{author}{Schnute, J.}, \bibinfo{year}{1985}.
\newblock \bibinfo{title}{A general theory for analysis of catch and effort
  data}.
\newblock \bibinfo{journal}{Canadian Journal of Fisheries and Aquatic Sciences}
  \bibinfo{volume}{42}, \bibinfo{pages}{414--429}.
\bibitem[{Somers and Wang(1997)}]{som97a}
\bibinfo{author}{Somers, I.}, \bibinfo{author}{Wang, Y.G.},
  \bibinfo{year}{1997}.
\newblock \bibinfo{title}{A simulation model for evaluating seasonal closures
  in australia's multispecies northern prawn fishery}.
\newblock \bibinfo{journal}{North American Journal of Fisheries Management}
  \bibinfo{volume}{17}, \bibinfo{pages}{114--130}.
\bibitem[{Sterling(2005a)}]{Sterling05r}
\bibinfo{author}{Sterling, D.}, \bibinfo{year}{2005}a.
\newblock \bibinfo{title}{Fishing energy efficiency review for the fisheries
  Research and development corporation}.
\newblock \bibinfo{type}{Technical Report} \bibinfo{number}{2005/239}.
  Fisheries Research and Development Corporation.
\bibitem[{Sterling(2005b)}]{sterlingPHD}
\bibinfo{author}{Sterling, D.}, \bibinfo{year}{2005}b.
\newblock \bibinfo{title}{\uppercase{m}odelling the physics of prawn trawling
  for fisheries management}.
\newblock Ph.D. thesis. School of applied physics.
  \bibinfo{address}{\uppercase{c}urtin \uppercase{u}niversity of
  \uppercase{t}echnology, \uppercase{p}erth, \uppercase{a}ustralia}.
\bibitem[{Walters(2003)}]{Wal03a}
\bibinfo{author}{Walters, C.}, \bibinfo{year}{2003}.
\newblock \bibinfo{title}{Folly and fantasy in the analysis of spatial catch
  rate data}.
\newblock \bibinfo{journal}{Canadian Journal of Fisheries and Aquatic Sciences}
  \bibinfo{volume}{60}, \bibinfo{pages}{1433--1436}.
\bibitem[{Wang(1999)}]{Wang99a}
\bibinfo{author}{Wang, Y.}, \bibinfo{year}{1999}.
\newblock \bibinfo{title}{A maximum-likelihood method for estimating natural
  mortality and catchability coefficient from catch and effort data}.
\newblock \bibinfo{journal}{Marine \& Freshwater Research}
  \bibinfo{volume}{50}, \bibinfo{pages}{307--11}.
\bibitem[{White(1975)}]{whit75a}
\bibinfo{author}{White, T.}, \bibinfo{year}{1975}.
\newblock \bibinfo{title}{Factors affecting the catchability of a penaeid
  shrimp {\it penaeus esculentus}.}, in: \bibinfo{editor}{Young, P.} (Ed.),
  \bibinfo{booktitle}{Proceedings of the First Australian National Prawn
  Seminar}, \bibinfo{publisher}{AGPS, Canberra, Australia}. pp.
  \bibinfo{pages}{115--137}.
\bibitem[{Zhou et~al.(2011)Zhou, Punt, Deng and
  Bishop}]{Zhoudoi:10.1139/f2011-052}
\bibinfo{author}{Zhou, S.}, \bibinfo{author}{Punt, A.E.},
  \bibinfo{author}{Deng, R.}, \bibinfo{author}{Bishop, J.},
  \bibinfo{year}{2011}.
\newblock \bibinfo{title}{Estimating multifleet catchability coefficients and
  natural mortality from fishery catch and effort data: comparison of bayesian
  state–space and observation error models}.
\newblock \bibinfo{journal}{Canadian Journal of Fisheries and Aquatic Sciences}
  \bibinfo{volume}{68}, \bibinfo{pages}{1171--1181}.

\end{thebibliography}
\end{document}